\newif\ifpdf
\ifx\pdfoutput\undefined
  \pdffalse  
\else
  \pdfoutput=1     
  \pdftrue
\fi

\documentclass{IEEEtran}
\usepackage[caption=false,font=footnotesize]{subfig}
\usepackage{fixltx2e}
\usepackage{amssymb}
\ifpdf
 \usepackage[pdftex]{graphicx}
\else
 \usepackage{graphicx}
\fi
\usepackage{url}
\usepackage{color}
\usepackage{setspace}
\usepackage{amsmath}
\usepackage{algorithm}
\usepackage{algpseudocode}
\usepackage{amsmath}
\usepackage{graphics}
\usepackage{epsfig}
\usepackage{cite}
\usepackage{array}
\newcolumntype{C}[1]{>{\centering\let\newline\\\arraybackslash\hspace{0pt}}m{#1}}
\newcolumntype{L}[1]{>{\raggedright\let\newline\\\arraybackslash\hspace{0pt}}m{#1}}
\newcolumntype{R}[1]{>{\raggedleft\let\newline\\\arraybackslash\hspace{0pt}}m{#1}}
\usepackage{multirow}

\usepackage{enumitem}

\newcommand{\defeq}{\stackrel{def}{=}}

\newcommand{\bm}[1]{\boldsymbol{#1}}
\newcommand\mb[1]{\mathbf{#1}}
\newcommand{\tr}{\mathrm{tr}}

\newcommand{\Diag}{\mathrm{Diag}}
\newcommand{\DIAG}{\mathrm{DIAG}}
\newcommand{\BER}{\mathrm{BER}}
\newcommand{\VEC}{\mathrm{vec}}

\newtheorem{lemma}{Lemma}

\newtheorem{remark}{Remark}

\begin{document}
\title{Detection of Spatially-Modulated Signals in Doubly Selective
Fading Channels With Imperfect CSI\thanks{This work was supported in part by Taiwan's Ministry of Science and Technology
under Grant NSC 99-2221-E-009-099-MY3. The material in this paper was presented in part at the 2013
IEEE Globecom Workshops.}
\author{Hsuan-Cheng Chang, Yen-Cheng Liu, and Yu T. Su$^\dag$\thanks{$\dag$Correspondence addressee.}\\}
\thanks{H.-C. Chang is with ASUSTeK Computer Inc., Taipei, Taiwan (email: Makoto\_Chang@asus.com).
Y.-C. Liu and Y. T. Su are with the Institute of Communications Engineering, National
Chiao Tung University, Hsinchu, Taiwan (email: ycliu@ieee.org; ytsu@nctu.edu.tw). }}

 \ifpdf
 \DeclareGraphicsExtensions{.pdf}
\else
 \DeclareGraphicsExtensions{.eps}
\fi

\maketitle \thispagestyle{empty}
\begin{abstract}
To detect spatially-modulated signals, a receiver needs the channel state information (CSI)
of each transmit-receive antenna pair. Although the CSI is never perfect and varies in time,
most studies on spatial modulation (SM) systems assume perfectly known CSI and time-invariant
channel. The spatial correlations among multiple spatial subchannels, which have to be considered
when CSI is imperfect, are also often neglected. In this paper, we release the above assumptions
and take the CSI uncertainty along with the spatial-temporal selectivities into account. We derive
the channel estimation error aware maximum likelihood (CEEA-ML) detectors as well as several
low-complexity alternatives for PSK and QAM signals. As the CSI uncertainty depends on the channel
estimator used, we consider both decision feedback and model based estimators in our study.

The error rate performance of the ML and some suboptimal detectors is analyzed. Numerical results
obtained by simulations and analysis show that the CEEA-ML detectors offer clear performance gain
against conventional mismatched SM detectors and, in many cases, the proposed suboptimal detectors
incur only minor performance loss.
\end{abstract}

\begin{IEEEkeywords}
  Imperfect channel state information, maximum likelihood, signal detection, space-time channel
  correlation, spatial modulation.
\end{IEEEkeywords}

\section{Introduction}
Spatial modulation (SM), as it allows only one transmit antenna to be active in any transmission
interval and exploits the transmit antenna index to carry extra information \cite{SM}, is a
low-complexity and spectral-efficient multi-antenna-based transmission scheme. Besides requiring
no multiple radio frequency (RF) transmit chains, its low complexity requirement is also due to
the fact that the receiver does not need complicated signal processing to deal with inter-spatial
channel interference (ICI).

Most receiver performance assessments on multi-antenna systems assume that the channel state
information (CSI) is perfectly known by the receiver (e.g., \cite{SM}). In practice, the CSI at
the receiver (CSIR) is obtained by a pilot-assisted or decision-directed (DD) estimator \cite{DD}
and is never perfect. The impacts of imperfect CSIR on some MIMO detectors were considered in
\cite{ImpCSI_MLdet1,MallikPerRx,ImpCSI_MLdet2} to evaluate the detectors' performance loss.
Furthermore, the channel is usually not static, especially in a mobile environment, but the
channel-aging effect, i.e., the impact of outdated CSI, is neglected in most of these studies
except for \cite{ImpCSI_MLdet2}. \cite{CSI_ph_err} discussed the effect of the channel coefficients'
{\it phase estimation error} while \cite{CE_eff2,CE_eff3} analyze the performance of the conventional
detector in the presence of channel estimation error which is assumed to be white Gaussian and
independent of the channel estimator used.
Other earlier works on SM detection all employ the perfect CSI and time-invariant channel
assumptions which thus yield suboptimal performance in practical environments \cite{CE_eff}.

The difficulty in deriving the optimal SM detector in a doubly (time and space) selective fading
channel with imperfect CSI is due to the fact that the likelihood function depends not only on the
transmitted symbol but also the CSI estimator and its performance. In this paper, we remove perfect
CSI and static channel assumptions and consider two MIMO channel estimators/trackers for systems
using a frame structure which inserts pilot symbols periodically among data blocks such that each
frame consists of a pilot block and several data blocks. The first scheme treats the detected data
as pilots to update channel estimate which is then used to detect the symbols of the following data
block. The second scheme models the channel variation by a polynomial so that channel estimation
becomes that of estimating the polynomial coefficients using three consecutive pilot blocks. We refer
to the first scheme as the decision-directed (DD) estimator (tracker) and the second one the model-based
(MB) estimator \cite{poly}.

Our {\it major contributions} are summarized as follows. We derive a general {\it channel estimation
error-aware} (CEEA) maximum likelihood (ML) receiver structure for detecting general MIMO signals with
practical channel estimators in doubly selective fading channels. The CEEA-ML detectors for $M$-PSK and
$M$-QAM based SM systems are obtained by specializing to the combined SM and PSK/QAM signal formats. As
the ML detectors require high computational complexity, we then develop two classes of low-complexity
suboptimal detectors. The first one detects the transmit antenna index and symbol separately (resulting
in two-stage detectors) while the second class simplifies the likelihood functions by using lower-dimension
approximations that neglect the spatial correlation on either the receive or transmit side. The approximations
lead to zero transmit correlation (ZTC) and zero receive correlation (ZRC) receivers. Both simplifications--separate
detection and dimension reduction of the likelihood function--can be combined to yield even simpler detector
structures. As will be verified later, the low-complexity detectors do not incur much performance loss. Except
for the two-stage detectors, we are able to analyze their error rate performance and confirm the accuracy of
the analysis by simulations. A model-based two-stage spatial correlation estimator is developed as well.

The rest of this paper is organized as follows. In Section \ref{sec:prelim} we present a general space-time
(S-T) MIMO channel model, review the corresponding perfect CSI based ML detector and introduce both DD and MB
channel estimators. We refer to the latter two channel estimators as DD-CE and MB-CE henceforth. In Section
\ref{sec:MLD}, we focus on MB-CE-aided systems. A general CEEA-ML detector for general MIMO signals is first
derived (cf. (\ref{eq:uni_ML_MB})), followed by those for $M$-PSK and $M$-QAM SM signals (cf. (\ref{eq:MBML_PSK})
and (\ref{eq:MBML_QAM})). A low-complexity two-stage receiver for $M$-PSK SM systems is given at the end of
the section (cf. (\ref{eq:MB_2Stage})). Section \ref{section:MLDD} presents similar derivations for DD-CE-aided
SM systems (cf. (\ref{eq:ML_DD}), (\ref{eq:DDML_PSK}), (\ref{eq:DDML_QAM}), and (\ref{eq:DD_2Stage})). In Section
\ref{sec:AML}, we develop simplified CEEA-ML and two-stage detectors for both MB- and DD-CE-aided SM systems by
using lower-dimension approximations of the likelihood functions. The error rate performance of various detectors
we derived is analyzed in Section \ref{sec:theo}. The analytic approach is valid for all but the two-stage detectors.
Because of space limitation, the presentation is concise, skipping most detailed derivations (cf. (\ref{eq:Approx_ST_MLMB}),
(\ref{eq:ZTC_MB}), (\ref{eq:MB_2SZRC}), (\ref{eq:ZRC_DD}), (\ref{eq:ZTC_DD}), and (\ref{eq:DD_2SZRC})). The
computational complexity and memory requirement of the mentioned detectors is analyzed in Section \ref{sec:complexity}.
As far as we know, materials presented in Sections \ref{sec:MLD}--\ref{sec:complexity} are new. Numerical performance
of our detectors and conventional mismatched detector is given in Section \ref{sec:sim}. Finally, we summarize our
main results and findings in Section \ref{sec:conclusion}.

\textit{Notations}: Upper and lower case bold symbols denote matrices and vectors, respectively. $\mb{I}_N$
is the $N\times N$ identity matrix and $\mb{0}_N$ the $N\times 1$ all-zero vector. $(\cdot)^T$, $(\cdot)^*$,
$(\cdot)^H$, $(\cdot)^{-1}$ and $(\cdot)^\dag$ represent the transpose, element-wise conjugate, conjugate
transpose, inverse, and pseudo-inverse of the enclosed items, respectively. $\mathrm{vec}(\cdot)$ is the operator
that forms one tall vector by stacking the columns of a matrix. $\mathbb{E}\{\cdot\}$, and $\|\cdot\|_{F}$ denote
the expectation and Frobenius norm of the enclosed items, respectively, $\otimes$ denotes the Kronecker product
and $\odot$ the Hadamard product. $(\cdot)_{i}$ and $[\cdot]_{ij}$ denote the $i$th row and $(i,j)$th element of
the enclosed matrix, respectively. $\Diag(\cdot)$ translates the enclosed vector or elements into a diagonal matrix
with the nonzero terms being the enclosed items, whereas $\DIAG(\cdot)$ is defined by
\begin{IEEEeqnarray}{rCl}
  \DIAG(\mb{x}_1,\mb{x}_2,\cdots,\mb{x}_M)
  =\left[
     \begin{array}{cccc}
       \mb{x}_1 & \mb{0}_{N_1} & \cdots & \mb{0}_{N_1} \\
       \mb{0}_{N_2} & \mb{x}_2 & \cdots & \mb{0}_{N_2} \\
       \vdots & \vdots & \ddots & \vdots \\
       \mb{0}_{N_M} & \mb{0}_{N_M} & \cdots & \mb{x}_M \\
     \end{array}
   \right]\notag
\end{IEEEeqnarray}
with vector length of $\mb{x}_i$ being $N_i$.

The detector structures we develop have to compute a common quadratic form and involve various conditional
mean vectors and covariance matrices as a function of the data block index $k$. For notational brevity and
the convenience of reference, we list the latter conditional parameters in Tables \ref{tab:notation_mb} and
\ref{tab:notation_dd} and define
\begin{equation}
  \mathcal{G}(\mb{\Gamma},\bm{\chi})\defeq\bm{\chi}^H\mb{\Gamma}^{-1}\bm{\chi}
\label{eq:G func}
\end{equation}
where $\mb\Gamma\in\mathbb{C}^{N_1\times N_1}$ is invertible and $\bm\chi\in\mathbb{C}^{N_1\times N_2}$.
\renewcommand{\arraystretch}{1.3}
\begin{table}[t]
\caption{Statistics Used in MB-CE-aided detectors, where subscript SSK stands for space-shift keying \cite{CSI_ph_err}}
 \centering
 \tabcolsep 0.02in
 \begin{tabular}{|C{.6in}C{.6in}|C{.9in}|C{.9in}|}
  \hline %
  \multicolumn{2}{|c|}{Conditional}& Mean & Covariance \\ \hline
  \multicolumn{1}{|c|}{\multirow{3}{*}{CEEA-ML}} & General & ${\mb{m}}_{mb}(k)$ (\ref{eq:m_mb}) &  ${\mb{C}}_{mb}(k)$ (\ref{eq:C_mb})\\ \cline{2-4}
  \multicolumn{1}{|c|}{} & PSK-SM & $\bar{\mb{m}}_{ssk}(k)$ & $\bar{\mb{C}}_{psk}(k)$ \\ \cline{2-4}
  \multicolumn{1}{|c|}{} & QAM-SM &  $\bar{\mb{m}}_{ssk}(k)$ & $\bar{\mb{C}}_{qam}(k)$ (\ref{eq:Cmb_qam}) \\ \hline
  \multicolumn{2}{|c|}{Two-stage} & $\bar{\mb{m}}_{ssk}(k)$ & $\bar{\mb{C}}_{psk}(k)$ \\\hline
  \multicolumn{2}{|c|}{ZRC} & $\bar{\mb{M}}_{zrc}(k)$ & $\bar{\mb{C}}_{zrc}(k)$ (\ref{eq:ApproxMLMBCov})\\ 
\hline
  \multicolumn{2}{|c|}{Two-stage ZRC} & $\bar{\mb{M}}_{ssk}(k)$ & $\bar{\mb{C}}_{ssk}(k)$ (\ref{eq:Cmb_zrc2s}) \\\hline
  \multicolumn{2}{|c|}{ZTC} & $\bar{\mb{M}}_{ztc}(k)$ (\ref{eq:Mmb_ztc}) & $\bar{\mb{C}}_{ztc}(k)$ (\ref{eq:Cmb_ztc})\\  \hline \end{tabular}%
 \label{tab:notation_mb}
\end{table}
\renewcommand{\arraystretch}{1}

\renewcommand{\arraystretch}{1.3}
\begin{table}[t]
\caption{Statistics Used in DD-CE-aided detectors}
 \centering
 \tabcolsep 0.02in
 \begin{tabular}{|C{.6in}C{.6in}|C{.9in}|C{.9in}|}
  \hline %
  \multicolumn{2}{|c|}{Conditional} & Mean & Covariance \\ \hline
  \multicolumn{1}{|c|}{\multirow{3}{*}{CEEA-ML}} & General & ${\mb{m}}_{dd}$ (\ref{eq:DDMLMean}) &  ${\mb{C}}_{dd}$ (\ref{eq:DDMLCov})\\ \cline{2-4}
  \multicolumn{1}{|c|}{} & PSK-SM & $\tilde{\mb{m}}_{ssk}$ & $\tilde{\mb{C}}_{psk}$ \\ \cline{2-4}
  \multicolumn{1}{|c|}{} & QAM-SM &  $\tilde{\mb{m}}_{ssk}$ & $\tilde{\mb{C}}_{qam}$  \\ \hline
  \multicolumn{2}{|c|}{Two-stage} & $\tilde{\mb{m}}_{ssk}$ & $\tilde{\mb{C}}_{psk}$ \\\hline
  \multicolumn{2}{|c|}{ZRC} & $\tilde{\mb{M}}_{zrc}$ & $\tilde{\mb{C}}_{zrc}$ (\ref{eq:Cdd_ZRC})\\ 
\hline
  \multicolumn{2}{|c|}{Two-stage ZRC} & $\tilde{\mb{M}}_{ssk}$ (\ref{eq:Mdd_ZRC2S}) & $\tilde{\mb{C}}_{ssk}$ (\ref{eq:Cdd_ZRC2S}) \\\hline
  \multicolumn{2}{|c|}{ZTC} & $\tilde{\mb{M}}_{ztc}$ (\ref{eq:Mdd_ZTC}) & $\tilde{\mb{C}}_{ztc}$ (\ref{eq:Cdd_ZTC})\\  \hline \end{tabular}%
 \label{tab:notation_dd}
\end{table}
\renewcommand{\arraystretch}{1}
\section{Preliminaries}\label{sec:prelim}

\subsection{S-T Correlated Channels}\label{section:ConventionMIMO}
We consider a MIMO system with $N_T$ transmit and $N_R$ receive
antennas and assume a block-faded scenario in which the MIMO channel
remains static within a block of $B$ channel uses but varies from
block to block. For this system, the received sample matrix of block $k$ can be
expressed as
\begin{IEEEeqnarray}{rCl}
   \mb{Y}(k)&\defeq&\left[\mb{y}_1(k),\cdots,\mb{y}_{B}(k)\right]
   \defeq\left[\underline{\mb{y}}_1^T(k),\cdots,\underline{\mb{y}}_{M}^T(k)\right]^T\notag\\
   &=&\mb{H}(k)\mb{X}(k)+\mb{Z}(k),
 \label{eq:ini_eq}
\end{IEEEeqnarray}
where $\mb{y}_i(k)$ are column vectors, $\underline{\mb{y}}_i(k)$ are row vectors, and
\begin{subequations}
\begin{IEEEeqnarray}{rCl}
  \mb{H}(k)\defeq[h_{ij}(k)]&=&[\mb{h}_1(k),\cdots,\mb{h}_{N_T}(k)]
\label{eq:H_ColForm}\\
&=&[\underline{\mb{h}}_1^T(k),\cdots,\underline{\mb{h}}_{N_R}^T(k)]^T
\label{eq:H_RowForm}
\end{IEEEeqnarray}
\end{subequations}
is the $N_R\times N_T$ channel matrix and
\begin{IEEEeqnarray}{rCl}
  \mb{X}(k)=[\mb{x}_1(k),\cdots,\mb{x}_{B}(k)]
=[\underline{\mb{x}}_1^T(k),\cdots,\underline{\mb{x}}_{N_T}^T(k)]^T
\end{IEEEeqnarray}
is the $N_{T}\times B$ ($B\geq N_T$) matrix containing the modulated data
or pilot symbols and the entries of the noise matrix $\mb{Z}(k)$ are i.i.d.
$\mathcal{CN}(0,\sigma^{2}_{z})$.

Let
$\mb{\Phi}=\mathbb{E}\left\{
     \mathrm{vec}\left(\mb{H}(k)\right)
     \mathrm{vec}^{H}\left(\mb{H}(k)\right)
     \right\}$
be the matrix whose $(N_{R}N_{T})^2$ entries represent the correlation
coefficients amongst spatial subchannels.
\begin{IEEEeqnarray}{rCl}
\mathrm{vec}\left(\mb{H}(k)\right)=\mb{\Phi}^{\frac{1}{2}}
 \mathrm{vec}\left(\mb{H}_{w}(k)\right),
\end{IEEEeqnarray}
where $\mb{H}_{w}(k)$ is an $N_{R}\times N_{T}$ matrix with i.i.d.
zero-mean, unit-variance complex Gaussian random variables as its elements.
We assume that the spatial correlation matrix $\mb{\Phi}$ is either completely
or partial known to the receiver in our derivations of various detector
structures. The effect of using estimated $\mb{\Phi}$ is studied by computer
simulation in Section \ref{sec:sim}.

We further assume that the spatial and temporal correlations are separable
\cite{ST_CorrCh}, i.e.,
\begin{IEEEeqnarray}{rCl}
\mathbb{E}\{h_{ij}(k)h_{mn}^{*}(\ell)\}=\rho_{S}(i-m,j-n)\times\rho_{T}(k-\ell)\notag
\end{IEEEeqnarray}
with $\rho_{T}(k-\ell)\equiv\mathbb{E}\{h_{ij}(k)h_{ij}^{*}(\ell)\}$
denoting the $(i,j)$th subchannel autocorrelation in time while
\begin{IEEEeqnarray}{rCl}
  \rho_{S}(i-m,j-n;k)&\equiv&\mathbb{E}\{h_{ij}(k)h_{mn}^{*}(k)\}\notag\\
&=&\big[\mb{\Phi}\big]_{(j-1)N_R+i,\:(n-1)N_R+m}
\label{non_Kron_rhoS}
\end{IEEEeqnarray}
the spatial correlation. As mentioned before, conditioned on the estimated channel matrix,
the likelihood function is a function of these S-T correlations and so are the corresponding
CEEA-ML detectors.

\subsection{Detection of SM Signals with Perfect CSIR}
\label{section:SMModel}
Although a spatial multiplexing (SMX) system such as BLAST \cite{BLAST} has a multiplexing gain,
it requires high-complexity signal detection algorithms to suppress ICI. The SM scheme avoids the
ICI-related problems by imposing the single active antenna constraint and compensates for the
corresponding data rate reduction by using the transmit antenna index to carry extra information
bits \cite{SM}.

An $m$-bit/transmission SM system partitions the data stream into groups of $m=\log_{2}(MN_{T})$
bits of which the first $\log_{2}N_{T}$ bits are used to determine the transmit antenna and the
remaining bits are mapped into a symbol in the constellation $\mathcal{A}_{M}$ of size $M$ for the
selected antenna to transmit. Since only one transmit antenna is active in each transmission, the
$j$th column of $\mb{X}(k)$ is of the form
\begin{equation}
  \mb{x}_j(k)=[0,\cdots,0,x_{\ell_j}(k),0,\cdots,0]^T,
  \label{eq:x_j}
\end{equation}
where $\ell_j\equiv\ell_j(k)$ is the active antenna index and $x_{\ell_j}(k)\defeq
s_j(k)\in \mathcal{A}_M$ is the modulated symbol transmitted at the $j$th symbol interval
of the $k$th block. A transmitted symbol block $\mb{X}$ can thus be decomposed as
$\mb{X}=\mb{L}\mb{S}$, where ${\mb{S}}=\Diag(\mb{s})$, ${\mb{s}}=[{s}_{1},\cdots,{s}_{B}]^T$,
$s_j \in \mathcal{A}_M$ and ${\mb{L}}$ is the $N_T\times B$ space-shift keying (SSK) matrix \cite{CE_eff2} defined as
\begin{IEEEeqnarray}{rCl}
{[{\mb{L}}]_{ij}}=\left\{
\begin{array}{ll}
  1, & \hbox{if $i=\ell_j$;} \\
  0, & \hbox{otherwise}.
\end{array}
\right.
\label{eq:L_mtx_def}
\end{IEEEeqnarray}
We define $\mathcal{X}$ as the set of all $N_T\times B$
matrices whose columns satisfy (\ref{eq:x_j}).
The average power
\begin{IEEEeqnarray}{rCl}\label{eq:avg_sigPwr}
    \varepsilon_s\defeq\frac{1}{B}\mathbb{E}\left\{
                \|\mb{X}(k)\|_F^{2}\right\}
            =\frac{1}{B}\mathbb{E}\left\{\tr\left(
            \mb{X}(k)\mb{X}^{H}(k)
            \right)\right\}\label{eq:ave_power}
\end{IEEEeqnarray}
is equivalent to the average power of $\mathcal{A}_M$.
Therefore, for $j=1,\cdots,B$,
the $j$th column vector in (\ref{eq:ini_eq}) can be written as
\begin{IEEEeqnarray}{rCl}\label{SMchannel}
\mb{y}_j(k)=\mb{h}_{\ell_j}(k) 
s_j(k)+\mb{z}_j(k).
\end{IEEEeqnarray}

\renewcommand{\arraystretch}{1}

Assuming perfect CSIR, i.e., $\hat{\mb{H}}(k)=\mb{H}(k)$, and i.i.d. source, we have the ML detector
\begin{IEEEeqnarray}{rCl}
\hat{\mb X}(k)
&=& \arg\underset{{\mb{X}}\in\mathcal{X}}{\max}\: P\Big(\mb{Y}(k)|\mb{H}(k),\mb{X}\Big)
\label{eq:optMis}
\end{IEEEeqnarray}
where
\begin{IEEEeqnarray}{rCl}
 &&P\Big(\mb{Y}(k)|\mb{H}(k),\mb{X}(k)\Big)\notag\\
 &&\hspace{2em}=\left(\pi\sigma_{z}^2\right)^{-N_{R}}
 \exp\left(-\frac{1}{\sigma_{z}^2}\left\|\mb{Y}(k)-\mb{H}(k)\mb{X}(k)\right\|_{F}^2\right).~~~
\notag
\end{IEEEeqnarray}
With $\mathbb{L}\defeq\{1,\cdots,N_T\}$, (\ref{eq:optMis}) is simplified as
\begin{IEEEeqnarray}{rCl}
\left(\hat{s}_j(k),\hat{\ell}_j(k)\right)
&=&\arg\underset{({s},{\ell})\in\mathcal{A}_M\times\mathbb{L}}{\min}\|
\mb{y}_{j}(k)-\mb{h}_{{\ell}}{s}\|_F^2
\end{IEEEeqnarray}
for $j=1,\cdots,B$. 

\subsection{Tracking Time-Varying MIMO Channels}
\label{sec:ChEst}
\begin{figure} \centering
\includegraphics[width=3.5in]{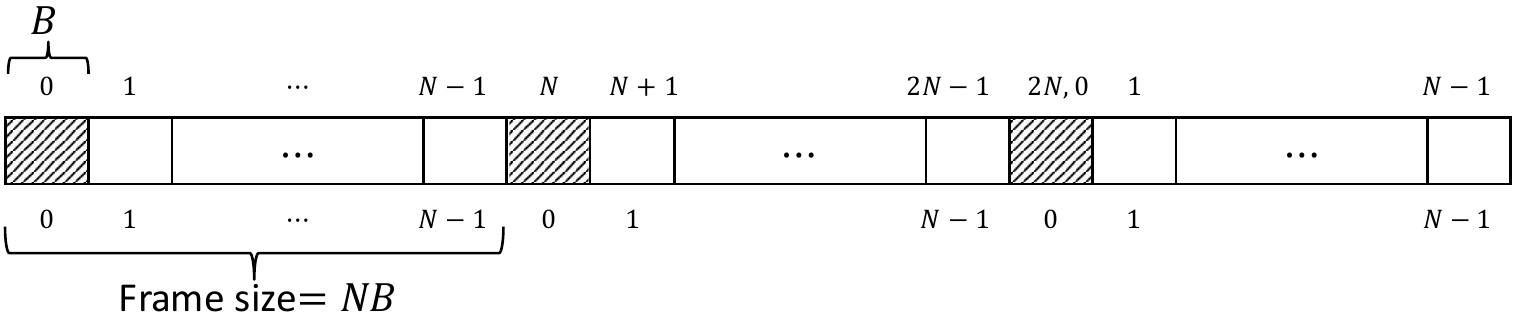}
\caption{Proposed frame structure. A pilot block, shown as a shaded area, is inserted every $N$ blocks. The block indices for MB-CE-aided systems ($0\leq k\leq2N$) and those for DD counterparts  ($0\leq k\leq N-1$) are labeled on the top and bottom of the blocks, respectively. 
}\label{fig:pilot}
\end{figure}

We now describe the two simple pilot-assisted channel trackers (estimators) to be considered in
subsequent discourse. The first one describes the channel's time variation by a polynomial while
the second one is a DD estimator that treats the detected data symbols as pilots. The frame
structure considered is depicted in Fig. \ref{fig:pilot}, where a frame consists of a pilot
block and $N-1$ data blocks with a total length of $NB$ symbol intervals. The choice of $N$
depends on the channel's coherence time \cite{pilot_struct}. For the MB-CE based system, we use three consecutive pilot
blocks to derive the CSI for $2N+1$ consecutive blocks and detect the associated data symbols.
The DD-CE uses the pilot block at the beginning of a frame to obtain an estimate of the S-T channel for detecting
the data of the next data block. This newly detected data block is then treated as pilots (called pseudo-pilots) to update
the channel information for detecting the ensuing data block. This pseudo-pilot-based channel estimation-data
detection procedure repeats until the last data block of the frame is detected and a new pilot block arrives.

Let the pilot block be transmitted at the $k_p$th block and denote the $N_T\times B$ pilot matrix
by $\mb{X}(k_p)=\mb{X}_p$. The corresponding received block is
\begin{IEEEeqnarray}{rCl}
    \mb{Y}(k_p)=\mb{H}(k_p)\mb{X}_p+\mb{Z}(k_p)
\end{IEEEeqnarray}
with the average pilot symbol power given by
$    \varepsilon_p\defeq\frac{1}{B}\|\mb{X}_p\|_F^{2}$
and
$\mb{X}_p$ a unitary matrix. In particular, for SM systems, we assume
that $B=N_T$ and $\mb{X}_p=\sqrt{\varepsilon_p}\mb{I}_{N_T}$.

\subsubsection{MB Channel Estimator (MB-CE)}
\label{section:EstMB}
For a single link with moderate mobility and frame size, it is reasonable to assume that
the $(i,j)$th component of the channel matrix ${\mb H}(k)$
is a quadratic function the sampling epoch, $\{k\}$ \cite{poly}
\begin{IEEEeqnarray}{rCl}
      h_{ij}(k)=\alpha_{ij}(k)k^{2}+\beta_{ij}(k)k+\gamma_{ij}(k).
\label{eq:MB_eq}
\end{IEEEeqnarray}
Define the coefficient vector $\bm{\xi}_{ij}(k)\defeq[\alpha_{ij}(k), \beta_{ij}(k),$
$\gamma_{ij}(k)]^T$. 
By collecting the received samples at three consecutive pilot locations $\mb{Y}(k_p)$,
$\mb{Y}(k_p+N)$,
$\mb{Y}(k_p+2N)$, we update the estimate for $\bm{\xi}_{ij}(k)$ every two frames via
\begin{IEEEeqnarray}{rCl}\label{eq:coef_eq}
\hat{\bm{\xi}}_{ij}(k_p)\defeq\left[%
 \begin{array}{c}
           \hat{\alpha}_{ij}(k_p)\\
           \hat{\beta}_{ij}(k_p)\\
           \hat{\gamma}_{ij}(k_p)\\
 \end{array}
 \right]
 =\mb{T}^{-1}(k_p)\:\tilde{\mb{y}}_{ij}(k_p)
\end{IEEEeqnarray}
where
\begin{subequations}
\begin{IEEEeqnarray}{rCl}
   \mb{T}(k)&\defeq&\left[%
 \begin{array}{ccc}
           k^2 & k & 1\\
           (k+N)^{2}&k+N& 1\\
           (k+2N)^{2}&k+2N& 1\\
    \end{array}
     \right],\\
%
\tilde{\mb{y}}_{ij}(k_p)&\defeq&\frac{1}{\sqrt{\varepsilon_p}}
\left[
 \begin{array}{c}
           y_{ij}(k_p)\\
           y_{ij}(k_p+N)\\
           y_{ij}(k_p+2N)\\
 \end{array}
 \right]\notag\\&=&
 \left[
 \begin{array}{c}
           h_{ij}(k_p)\\
           h_{ij}(k_p+N)\\
           h_{ij}(k_p+2N)\\
 \end{array}
 \right]
 +\frac{1}{\sqrt{\varepsilon_p}}\tilde{\mb{z}}_{ij}(k_p),
\label{eq:pilot_eq}
\end{IEEEeqnarray}
\end{subequations}
and $\tilde{\mb{z}}_{ij}(k_p)\sim\mathcal{CN}(\mb{0}_3,\frac{\sigma_z^2}{{\varepsilon_p}}\mb{I}_3)$.
From (\ref{eq:MB_eq}), (\ref{eq:coef_eq}) and define
$\mb{t}(k)=[k^2,k,1]^T$, we obtain the MB-CEs
 \begin{IEEEeqnarray}{rCl}
  \hat{h}_{ij}(k)=\mb{t}^H(k)\hat{\bm{\xi}}_{ij}(k_p)
 =\mb{t}^H(k)\:\mb{T}^{-1}(k_p)\:\tilde{\mb{y}}_{ij}(k_p)
\label{eq:MB_Hhat}
\end{IEEEeqnarray}
for channels at blocks $k_p,\cdots,k_p+2N-1$.

\subsubsection{DD Channel Estimator (DD-CE)}
\label{section:EstDD}
The DD-CE uses the detected data symbols in a data block as pilots to obtain an updated CE (estimated CSI)
and then apply it for detecting the data symbols in the next data block. Error propagation, if exists, is
terminated at the end of a frame as the new frame uses the new pilot block $\mb{X}(k)$ to obtain a new CE, e.g.,
the least squares (LS) estimate $\hat{\mb{H}}(k_p+sN) \stackrel{def}{=}\hat{\mb{H}}(k)=$ $\mb{Y}(k)
\hat{\mb{X}}^{\dagger}(k)$, $k=k_p+sN, s\in\mathbb{Z}$.

We note that only one element in each column of the $k$th detected block $\hat{\mb{X}}(k)$ is nonzero, hence
it is likely that not all vectors of CEs $\hat{\mb h}_\ell(k)$'s are updated at each data block and $\mb{X}(k)$
or $\hat{\mb{X}}(k)$ is not of full rank most of the time. However, in the long run, all channel coefficients
would be updated as each transmit antenna is equally likely selected.
We denote by $\hat{\mb{H}}[\mathfrak{L}](k)$, the submatrix containing only the columns
associated with the set of antenna indices activated in block $k$, $\mathfrak{L}\subseteq
\{1,\cdots,N_T\}$, i.e.,
\begin{IEEEeqnarray}{rCl}\label{DD1}
  \hat{\mb{H}}[\hat{\mathfrak L}(k)](k)\leftarrow
\mb{Y}(k)\bar{\mb{X}}^{\dagger}(k),\notag
\end{IEEEeqnarray}
where $\hat{\mathfrak L}(k)\defeq\{\hat{\ell}_{1}(k),\cdots, \hat{\ell}_{B}(k)\}$ and
$\bar{\mb{X}}(k)$ is the truncated $\hat{\mb{X}}(k)$ with its all-zero rows removed.
By combining the submatrix which consists of those channel vectors estimated in the previous
(the $(k-1)$th) block
\begin{IEEEeqnarray}{rCl}\label{DD2}
    \hat{\mb{H}}[\mathbb{L} \setminus \hat{\mathfrak L}(k)](k)
    =\hat{\mb{H}}[\mathbb{L} \setminus \hat{\mathfrak L}(k)](k-1),\notag
\end{IEEEeqnarray}
we obtain a full-rank channel matrix estimate $\hat{\mb{H}}(k)$ and, with a slight abuse of notation,
continue to denote $\mb{Y}(k)\hat{\mb{X}}^{\dagger}(k)$ as $\hat{\mb{H}}(k)$.
\begin{remark}
As the DD method uses the CE obtained in the previous block for demodulating the current block's data,
error propagation within a frame is inevitable. The MB approach avoids error propagation at the cost
of increased latency and storage requirement. Detailed computing complexity and memory requirement are
given in Section \ref{sec:complexity}.

The DD method updates the CSI estimate in each block while the channel's time variation is taken into
account by the MB approach through the model (\ref{eq:MB_eq}) which leads to the estimate (\ref{eq:MB_Hhat})
in an $N$-block frame. $\square$
\end{remark}

In the remainder of this paper, we assume the normalization $\varepsilon_p=1$ and in Section
\ref{sec:sim}, $\varepsilon_p=\varepsilon_s=1$.

\section{Channel Estimation Error-Aware ML Detection With MB Channel Estimates}\label{sec:MLD}
As defined in \cite{ImpCSI_MLdet1}, a {\it mismatched detector} is the one which replaces ${\mb H}(k)$
in (\ref{eq:optMis}) by the estimated CSI $\hat{\mb{H}}$
\begin{IEEEeqnarray}{rCl}
 \hat{\mb{X}}^\text{MM}(k)
  &\defeq&\arg \underset{\mb{X}\in\mathcal{X}}
  {\min}
    \|\mb{Y}(k)-\hat{\mb{H}}\mb{X}\|_F^2.
  \label{eq:MM_D}
\end{IEEEeqnarray}
In our case, $\hat{\mb{H}}$ is either
$\hat{\mb{H}}(k)$ or $\hat{\mb{H}}(k-1)$ depending on whether an MB or DD estimator is used.

We extend the basic approach of \cite{ImpCSI_MLdet1} by taking the channel aging effect and the
spatial correlation into account and refer to the resulting detectors as {\it channel
estimation error-aware (CEEA)-ML detectors}. While most of the existing works assume a
time-invariant environment, we assume that the channel varies from block to block and is
spatial/temporally correlated. As in \cite{ImpCSI_MLdet1} we need the following lemma in
subseuqent analysis.
\begin{lemma}
\cite[Thm. 10.2]{MLEstimation}
Let $\mb{z}_{1}$ and $\mb{z}_{2}$ be circularly symmetric complex Gaussian random vectors with
zero means and full-rank covariance matrices $\mb{\Sigma}_{ij}\defeq\mathbb{E}\{\mb{z}_{i}
\mb{z}_{j}^{H}\}$. Then, conditioned on $\mb{z}_{2}$, the random vector $\mb{z}_{1}$ is circularly
symmetric Gaussian with mean $\mb{\Sigma}_{12}\mb{\Sigma}_{22}^{-1}\mb{z}_{2}$ and covariance matrix
$\mb{\Sigma}_{11}-\mb{\Sigma}_{12}\mb{\Sigma}_{22}^{-1} \mb{\Sigma}_{21}$. \label{lemma}
\end{lemma}

\subsection{General MIMO Signal Detection with Imperfect CSI}\label{section:exactML}
When the MB-CE $\hat{\mb{H}}(k)$ is used, the {\it CEEA-ML} MIMO detector
has to compute
\begin{IEEEeqnarray}{rCl}
   \arg\underset{\mb{X}\in\mathcal{A}_{M^+}^{N_T\times B}}{\max}
   \:P\Big[\mathrm{vec}(\mb{Y}(k))\Big|
    \mathrm{vec}(\mb{X}),\mathrm{vec}(\hat{\mb{H}}(k))\Big],~~~
\label{eq:MBlikelihood}
\end{IEEEeqnarray}
where $\mathcal{A}_{M^+}\defeq\mathcal{A}_M\cup\{0\}$.
Since all the entries of $\mb{Y}(k)$ and $\hat{\mb{H}}(k)$
are zero-mean random variables, we invoke \textit{Lemma \ref{lemma}} with
\begin{subequations}
\begin{IEEEeqnarray}{rCl}
\mb{z}_{1}&=&\mathrm{vec}(\mb{Y}(k))\notag\\
&=&\left(\mb{X}^{T}(k)\otimes\mb{I}_{N_R}\right)\mathrm{vec}(\mb{H}(k))
 +\mathrm{vec}(\mb{Z}(k)),\label{eq:MBCE_z1}\\
\mb{z}_{2}&=&\mathrm{vec}(\hat{\mb{H}}(k))\notag\\
&=&\text{\small{$\Big[\mathrm{vec}(\mb{Y}(k_p))~\mathrm{vec}(\mb{Y}(k_p+N))
 ~\mathrm{vec}(\mb{Y}(k_p+2N))\Big]$}}\notag\\
 &&\times \left(\mb{t}^H(k)\mb{T}^{-1}(k_{p})\right)^{T},
 \label{eq:MBCE_z2}%
 \end{IEEEeqnarray}
\end{subequations}
for $k_p<k<k_p+2N$, to obtain
\begin{subequations}
\begin{IEEEeqnarray}{rCl}
\mb{\Sigma}_{11}&=&\mathbb{E}\{\mb{z}_{1}\mb{z}_{1}^{H}\}\notag\\
&=&\left(\mb{X}^{T}(k)\otimes\mb{I}_{N_R}\right)
\mb{\Phi}(\mb{X}^{\ast}(k)\otimes\mb{I}_{N_R})
+\sigma^{2}_{z}\mb{I}_{N_RN_T},~~~~~\label{eq:MBMLSigma11}\\
\mb{\Sigma}_{12}&=&\mathbb{E}\{\mb{z}_{1}\mb{z}_{2}^{H}\}=
\mathbb{E}\{(\mb{z}_{2}\mb{z}_{1}^{H})^{H}\}=\mb{\Sigma}^{H}_{21}\notag\\
&=&\mb{t}^H(k)\mb{T}^{-1}(k_{p})\mb{q}(k)
\left(\mb{X}^{T}(k)\otimes\mb{I}_{N_R}\right)\mb{\Phi},\label{eq:MBMLSigma12}\\
\mb{\Sigma}_{22}&=&\mathbb{E}\{\mb{z}_{2}\mb{z}_{2}^{H}\}\notag\\
&=&\nu(k)\mb{\Phi}
+\sigma^{2}_{z}\left\|\mb{t}^H(k)\mb{T}^{-1}(k_{p})\right\|_F^{2}
\mb{I}_{N_RN_T},\label{eq:MBMLSigma22}%
\end{IEEEeqnarray}
\end{subequations}
where
\begin{subequations}
\begin{IEEEeqnarray}{rCl}
\mb{q}(k)&=&
    \Big[\rho_{T}(k-k_p),\rho_{T}(k-k_p-N),\rho_{T}(k-k_p-2N)\Big]^T,\notag\\ \label{eq:q}\\
\nu(k)&=&\mb{t}^H(k)\mb{T}^{-1}(k_{p})
\left[
\begin{array}{ccc}
 1& \rho_{T}(N)& \rho_{T}(2N)\\
 \rho_{T}(N)& 1& \rho_{T}(N)\\
 \rho_{T}(2N)& \rho_{T}(N)& 1\\
 \end{array}
\right]\notag\\
&&\times \left(\mb{t}^H(k)\mb{T}^{-1}(k_{p})\right)^{H}.\label{eq:nu}
\end{IEEEeqnarray}
\label{eq:q_nu}
\end{subequations}

Direct substitutions of the above covariance matrices, we obtain
the mean vector ${\mb{m}}_{mb}(k)$ and covariance matrix
$\mb{C}_{mb}(k)$
\begin{subequations}
\begin{IEEEeqnarray}{rCl}
\label{eq:MBMLMean}
{\mb{m}}_{mb}(k)
&=&\left(\mb{X}^{T}(k)\otimes\mb{I}_{N_R}\right)\mb{A}(k)
\mathrm{vec}(\hat{\mb{H}}(k)),\label{eq:m_mb}\\
\mb{C}_{mb}(k)
&=&\sigma^{2}_{z}\mb{I}_{N_RN_T}
+\left(\mb{X}^{T}(k)\otimes\mb{I}_{N_R}\right)\Big[\mb{I}_{N_RN_T}-\mb{A}(k)~~~
\notag\\&&
\left(\mb{t}^H(k)\mb{T}^{-1}(k_{p})\mb{q}(k)\right)^*\Big]
\mb{\Phi}
\left(\mb{X}^{\ast}(k)\otimes\mb{I}_{N_R}\right)\label{eq:C_mb}
\end{IEEEeqnarray}
for the likelihood function of $\mb{Y}(k)$, where the spatial correlation
$\mb{\Phi}$ also appears in
\begin{equation}
  \mb{A}(k)=\mb{t}^H(k)
\mb{T}^{-1}(k_{p})\mb{q}(k)\:\mb{\Phi}\:\mb{\Sigma}_{22}^{-1}.
\label{eq:ML_MB_A}%
\end{equation}
\end{subequations}
Using (\ref{eq:G func}) and (\ref{eq:MBlikelihood})--(\ref{eq:ML_MB_A}), we obtain a more compact
form of the {\it CEEA-ML detector}
\begin{IEEEeqnarray}{rCl}\label{eq:MBML}
\hat{\mb{X}}^{\textrm{ML}}_{mb}(k)=\arg\underset{\mb{X}\in\mathcal{A}_{M^+}^{N_T\times B}}
{\min}&&\mathcal{G}(\mb{C}_{mb}(k),
\mathrm{vec}(\mb{Y}(k))-{\mb{m}}_{mb}(k))
\notag\\
&&+\log\det{{\mb{C}}_{mb}}(k).
\label{eq:uni_ML_MB}
\end{IEEEeqnarray}
An alternate derivation of $\mb{\Sigma}_{11}$ and $\mb{\Sigma}_{12}$ begins with
$\hat{\mb{H}}(k)=\mb{H}(k)+\mb{E}(k).\label{eq:CSI_err}$
It is verifiable that $\mathbb{E}\{\mathrm{vec}(\mb{E}(k))\}
 =\mb{0}_{N_RN_T}$ and
\begin{IEEEeqnarray}{rCl}
\mb{\Psi}_{E}(k)&\defeq&\mathbb{E}\Big\{\mathrm{vec}(\mb{E}(k))\mathrm{vec}^{H}(\mb{E}(k))\Big\}
 \notag\\
 &=&\Big(\nu(k)-2\mb{t}^H(k)\mb{T}^{-1}(k_{p})\mb{q}(k)
      +1\Big)\mb{\Phi}\notag\\
&& +\sigma^{2}_{z}\left\|\mb{t}^H(k)\mb{T}^{-1}(k_{p})\right\|_F^{2}
\mb{I}_{\tiny{N_RN_T}}.~~~~~~\label{eq:MBML_ErrMeanCov}
\end{IEEEeqnarray}
The received sample vector is thus related to $\mb{E}(k)$ via
\begin{IEEEeqnarray}{rCl}\label{eq:MBML_Y}
\mb{Y}(k)&=&\hat{\mb{H}}(k)\mb{X}(k)+\Big(\mb{Z}(k)-\mb{E}(k)\mb{X}(k)\Big)\notag\\
&=&\hat{\mb{H}}(k)\mb{X}(k)+\tilde{\mb{Z}}(k),
\end{IEEEeqnarray}
where $\mathrm{vec}(\tilde{\mb{Z}}(k))\sim \mathcal{CN}(\mb{0}_{N_RN_T},
\sigma^{2}_{z}\mb{I}_{N_RN_T}+(\mb{X}^{T}(k)\otimes\mb{I}_{N_R})$ $\mb{\Psi}_{E}(k)(\mb{X}^*(k)\otimes\mb{I}_{N_R}))$.
With
\begin{IEEEeqnarray}{rCl}
\mb{z}_1=\left(\mb{X}^{T}(k)\otimes\mb{I}_{N_R}\right)
\mathrm{vec}(\hat{\mb{H}}(k)) + \mathrm{vec}(\tilde{\mb{Z}}(k)),\notag
\end{IEEEeqnarray}
we obtain $\mb{\Sigma}_{11}$ and $\mb{\Sigma}_{12}$ as given by
(\ref{eq:MBMLSigma11}) and (\ref{eq:MBMLSigma12}).
%
\begin{remark}
A closer look at the components, (\ref{eq:MBCE_z1})--(\ref{eq:ML_MB_A}), of the {\it CEEA-ML detector}
(\ref{eq:uni_ML_MB}) reveals that the spatial correlation $\mb{\Phi}$ affects the auto-correlations
of the received signal and estimated channel, $\mb{\Sigma}_{11}$ and $\mb{\Sigma}_{22}$, and their
cross-correlations, $\mb{\Sigma}_{12}$ and $\mb{\Sigma}_{21}$. The influences of the time selectivity
$\rho_T(\cdot)$, frame structure (a pilot block in every $N$ blocks) and estimator structure (\ref{eq:MB_Hhat})
on the latter three correlations can be easily found in (\ref{eq:MBMLSigma12}) and (\ref{eq:MBMLSigma22})
and through $\mb{q}(k)$ and $\nu(k)$. These channel and system factors also appear in the estimator error
covariance $\mb{\Psi}_{E}(k)$.

In contrast, in analyzing the performance of {\it mismatched detectors} and the effects of imperfect CSI,
\cite{CE_eff,CE_eff2}, and \cite{CE_eff3} assume a simplified CSI error model that $\mb E(k)$ consists
only of white Gaussian components which are independent of the S-T correlations and channel estimator used,
although the latter is a critical and inseparable part of the detector structure. $\square$
\end{remark}

Note that (\ref{eq:uni_ML_MB}) is general enough to describe the detector structures for arbitrary data
format and/or modulation schemes. For SMX signals, the search range is modified to $\mathcal{A}_{M}^{N_T\times B}$,
while a precoded MIMO system, we replace $\mb X$ by $\mb W\mb S$ with $\mb W$ being the precoding matrix.
We derive specific detector structures for various SM signals in the remaining part of this section and
in Sections IV and V.

\subsection{CEEA-ML Detectors for SM Signals}
\label{section:MBCE_MLSM}
\subsubsection{$M$-PSK Constellation}
With the decomposition $\mb{X}=\mb{L}\mb{S}$ defined in Section \ref{section:SMModel} and $\mathcal{A}_M$
an $M$-PSK constellation, the {\it CEEA-ML detector} for the PSK-based SM system is
derivable from (\ref{eq:MBML}) and is given by
\begin{IEEEeqnarray}{rCl}
\hat{\mb{X}}^{\textrm{ML}}_{mb}(k)&=&
\arg\underset{(\mb{s},\mb{L})\in\mathcal{A}_M^B\times\mathcal{L}}
{\widetilde{\min}}\log\det\bar{\mb{C}}_{psk}(k)\notag\\
&&+\mathcal{G}(\bar{\mb{C}}_{psk}(k),{\mb{y}}_s(k)-\bar{\mb{m}}_{ssk}(k))
\label{eq:MBML_PSK}
\end{IEEEeqnarray}
where $\mathcal{L}$ denotes the set of all SSK matrices of the form
(\ref{eq:L_mtx_def}) $\bar{\mb{m}}_{ssk}(k)$ and $\bar{\mb{C}}_{psk}(k)$, which are not exactly the
conditional mean and covariance, are obtained from (\ref{eq:m_mb}) and (\ref{eq:C_mb})
by substituting $\mb{L}$ for $\mb{X}$ and $\frac{\sigma_{z}^{2}}{\varepsilon_s}$,
$\varepsilon_s=|s_j|^2$, for $\sigma_{z}^{2}$.
In (\ref{eq:MBML_PSK}), we have defined, for an implicit function $f$
of $\mb{s}$ and $\mb{L}$,
\begin{equation}
\arg\underset{(\mb{s},\mb{L})}{\widetilde{\min}}\:f(\mb{s},\mb{L})=
\hat{\mb{L}}~\Diag(\hat{\mb{s}}),~~~(\hat{\mb{s}},\hat{\mb{L}})=\arg\underset{(\mb{s},\mb{L})}
{{\min}}\:f(\mb{s},\mb{L}),
\end{equation}
where ${\mb{y}}_s(k)\defeq \mathrm{vec}\left(\mb{Y}(k){\mb{S}}^{H}\right)/{\varepsilon_s}$.

\subsubsection{$M$-QAM Constellation}
If $\mathcal{A}_M$ is an $M$-QAM constellation,
the corresponding {\it CEEA-ML detector} is
\begin{IEEEeqnarray}{rCl}
\hat{\mb{X}}^\text{ML}_{mb}(k)&=&
\arg\underset{(\mb{s},\mb{L})\in\mathcal{A}_M^B\times\mathcal{L}}{\widetilde{\min}}
N_R\log\det{\mb{E}}_s
+\log\det\bar{\mb{C}}_{qam}(k)\notag\\
&&+\mathcal{G}(\bar{\mb{C}}_{qam}(k),{\mb{y}}_s(k)-\bar{\mb{m}}_{ssk}(k))
\label{eq:MBML_QAM}
\end{IEEEeqnarray}
with ${\mb E}_s \defeq {\mb{S}}{\mb{S}}^H$, ${\mb{y}}_s(k)
\defeq \mathrm{vec}\left(\mb{Y}(k){\mb{S}}^H{\mb{E}}_s^{-1}\right)$, and
\begin{IEEEeqnarray}{rCl}
\bar{\mb{C}}_{qam}(k)&\defeq&
{\sigma^{2}_{z}}({\mb{E}}_s^{-1}\otimes\mb{I}_{N_R})
+({\mb{L}}^{T}\otimes\mb{I}_{N_R})\Big[\mb{I}_{N_RN_T}-
\mb{A}(k)\notag\\
& & \times\: \left(\mb{t}^H(k)\mb{T}^{-1}(k_{p})\mb{q}(k)\right)^*\Big]
\mb{\Phi}({\mb{L}}^{\ast}\otimes\mb{I}_{N_R}).~~~~~\label{eq:Cmb_qam}
\end{IEEEeqnarray}

\subsection{Two-Stage $M$-PSK SM Detector}
\label{subsec:CR_ML_MB}

The ML detector (\ref{eq:MBML_PSK}) calls for an exhaustive search over
the set of all candidate antenna index-modulated symbol pairs,
which has a cardinality of $|\mathbb{L}\times\mathcal{A}_M|^B$.
A low-complexity alternative which reduces the search dimension while
keeping the performance loss to a minimum would be desirable. Toward
this end, we consider a two-stage approach that detects the active antenna
indices and then the transmitted symbols in each block. A similar approach
with perfect CSIR assumption have been suggested in 
\cite{2Stage}.

We first notice that
\begin{IEEEeqnarray}{rCl}
&&\hspace{0em}P\Big[\mb{Y}(k)\Big|\mb{L}(k),\hat{\mb{H}}(k)\Big]\notag\\
&&\hspace{1em}=\sum_{s_{1}(k)}\sum_{s_{2}(k)}\cdots\sum_{s_{B}(k)}
P\Big[\mb{Y}(k)\Big|\mb{L}(k),\mb{S}(k),\hat{\mb{H}}(k)\Big]
P\Big[\mb{S}(k)\Big]\notag
\end{IEEEeqnarray}
where $\mb{L}(k)$ and $\mb{S}(k)$ are the $k$th block's SSK and symbol matrices.
It follows that the estimate of the activated antenna indices is
\begin{IEEEeqnarray}{rCl}\label{eq:SMML_PSK}
\hat{\mb{L}}(k)&=& \arg\underset{{\mb{L}}\in\mathcal{L}}{\max}~\frac{1}
{{M}^{B}\sqrt{\det \varepsilon_s\bar{\mb{C}}_{psk}(k)}}\sum_{{\mb{s}}\in\mathcal{A}_M^{B}}
\mathcal{F}({\mb s})
\end{IEEEeqnarray}
where
\begin{subequations}
\begin{IEEEeqnarray}{rCl}\label{Fs}
\mathcal{F}(\mb{s})&\defeq&\exp\left[\frac{-
\mathcal{G}(\bar{\mb{C}}_{psk}(k),{\bar{\mb{m}}_{ssk}}(k))}{2}
-\frac{{\mb{s}}^{T}{\mb{J}}(k){\mb{s}}^{*}}{2\varepsilon_s^{2}}\right.\nonumber\\
&&\left. ~~~~~~+\frac{\Re\{{\mb{b}}^H(k){\mb{s}}^{*}\}}{\varepsilon_s}\right],\\
{\mb{J}}(k)&=&\mathcal{G}\left(\bar{\mb{C}}_{psk}(k),\DIAG(\mb{y}_{1}(k),\cdots,\mb{y}_{B}(k))\right),
\label{eq:ML_2S_DD_J}\\
{\mb{b}}(k)&=&\DIAG\left(\mb{y}_{1}(k),\cdots,\mb{y}_{B}(k)\right)^H
\bar{\mb{C}}_{psk}^{-1}(k)\bar{\mb{m}}_{ssk}(k).\notag\\\label{eq:ML_2S_DD_b}
\end{IEEEeqnarray}
\end{subequations}
For each candidate SSK matrix ${\mb{L}}$, we have the approximation
\begin{IEEEeqnarray}{rCl}
\sum_{{\mb{s}}}\mathcal{F}({\mb s})
\approx \mathcal{F}(\bar{\mb s}(\mb{L})),~~\bar{\mb s}({\mb L})
\defeq\mathcal{Q}_{\mathcal{A}_{M}}\left({\tilde{\mb s}}({\mb L})\right),
\label{eq:MB_NML_App}
\end{IEEEeqnarray}
where $\mathcal{Q}_{\mathcal{A}_{M}}(\cdot)$ quantizing the enclosed items to the nearest
constellation points in $\mathcal{A}_{M}$ and ${\tilde{\mb s}}({\mb L})=\varepsilon_s({\mb{J}}^{-1}(k)
{\mb{b}}(k))^{*}$ being the solution of $\partial \mathcal{F}(\mb{s})/\partial\mb{s}=\mb{0}_B$.
$\mb{b}(k)$ defined in (\ref{eq:ML_2S_DD_b}) is an implicit function of $\mb{L}$ since as
mentioned in the previous subsection, $\bar{\mb{C}}_{psk}(k)$ and ${\bar{\mb{m}}_{ssk}}(k)$
are both functions of $\mb{L}$. The approximation (\ref{eq:MB_NML_App}) thus depends on
${\mb L}$ and the associated tentative demodulation decision $\bar{\mb s}({\mb L})$. A simpler
antenna indices estimate is then given by
\begin{subequations}
\begin{IEEEeqnarray}{rCl}
\hat{\mb{L}}(k)\approx \arg\underset{{\mb{L}}\in\mathcal{L}}{\min}\:&&
\log\det(\varepsilon_s\bar{\mb{C}}_{psk}(k))
+\mathcal{G}(\bar{\mb{C}}_{psk}(k),{\bar{\mb{m}}_{ssk}}(k))
\notag\\
&&\text{\small{$+\frac{\bar{\mb{s}}^{T}({\mb{L}})
{\mb{J}}(k)\bar{\mb{s}}^{*}({\mb{L}})}{\varepsilon_s^{2}}
-\frac{2\Re\left\{{\mb{b}}^H(k)\bar{\mb{s}}^{*}
({\mb{L}})\right\}}{\varepsilon_s}$}}.~~~~~
\label{eq:SMML_redu}
\end{IEEEeqnarray}
Once the active SSK matrix is determined, we output the decision $\bar{\mb s}({\mb L})$
associated with $\hat{\mb{L}}(k)$ which has been determined in the previous stage
\begin{IEEEeqnarray}{rCl}
\hat{\mb{s}}(k)=\bar{\mb{s}}(\hat{\mb{L}}(k)).
\label{eq:SMML_redu0}
\end{IEEEeqnarray}
\label{eq:MB_2Stage}%
\end{subequations}
(\ref{eq:SMML_redu0}) will reappear later repeatedly, each with different antenna index detection rule
$\hat{\mb{L}}(k)$. Obviously, the search complexity of the detector (\ref{eq:SMML_redu}) is much lower
than that of (\ref{eq:MBML_PSK}). For systems employing QAM constellations, the approximation (\ref{eq:MB_NML_App})
is not directly applicable as the nonconstant-modulus nature of QAM implies that $\mathcal{F}(\mb{s})$
has an additional amplitude-dependent term in the exponent.

\section{DD CE-Aided CEEA-ML Detectors}
\label{section:MLDD}
\subsection{General MIMO Signal Detection with Imperfect CSI}\label{subsec:GenMLDD}
To derive the {\it CEEA-ML detector} for general MIMO signals using a DD LS channel estimate, we again
appeal to \textit{Lemma \ref{lemma}} with $\mb{z}_{1}$ defined by (\ref{eq:MBCE_z1}) and $\mb{z}_{2}
\defeq\mathrm{vec}(\hat{\mb{H}}(k-1))=\mathrm{vec}(\mb{Y}(k-1)\hat{\mb{X}}^{\dag}(k-1))$. For $k_p <k
< k_p+N$, the signal block which maximizes the likelihood function $P\Big[\mathrm{vec}(\mb{Y}(k))
\Big|\mathrm{vec}(\mb{X}(k)),\mathrm{vec}(\hat{\mb{H}}(k-1))\Big]$ is given by
\begin{IEEEeqnarray}{rCl}
\hat{\mb{X}}^{\textrm{ML}}_{dd}(k)=\arg
\underset{\mb{X}\in\mathcal{A}_{M^+}^{N_T\times B}}{\min}\:&&
\log\det{\mb{C}}_{dd}\notag\\
&&+
\mathcal{G}({\mb{C}}_{dd},\mathrm{vec}(\mb{Y}(k))-{\mb{m}}_{dd}).~~~~~~
\label{eq:ML_DD}
\end{IEEEeqnarray}
To have more compact expressions we use $\hat{\mb{H}}$ and $\hat{\mb{H}}(k-1)$
interchangeably for the DD {\it CEEA-ML detectors}. This is justifiable as the conditional
mean and covariance of $\mb{Y}(k)$ given $\mb{X}(k)$ and $\hat{\mb{H}}(k-1)$ are
\begin{subequations}
\begin{IEEEeqnarray}{rCl}
{\mb{m}}_{dd}&=&
\left(\mb{X}^{T}(k)\otimes\mb{I}_{N_R}\right)\mb{A}
\mathrm{vec}(\hat{\mb{H}}(k-1)),\label{eq:DDMLMean}\\
 \mb{C}_{dd}&=&\sigma^{2}_{z}\mb{I}_{BN_R}
+\Big(\mb{X}^{T}(k)\otimes\mb{I}_{N_R}\Big)\Big[\mb{I}_{N_RN_T}-\rho_{T}(1)\mb{A}\Big]\notag\\
&&\cdot\:\mb{\Phi}\Big(\mb{X}^{\ast}(k)\otimes\mb{I}_{N_R}\Big)~~~~~
\label{eq:DDMLCov}
\end{IEEEeqnarray}
\label{eq:ML_DD_subs}
\end{subequations}
with
$  \mb{A}\defeq\rho_{T}(1)\:\mb{\Phi}\:
(\mb{\Phi}+\sigma_z^{2}\mb{I}_{N_RN_T})^{-1}$.

\subsection{CEEA-ML SM Signal Detectors}
Following the derivation of Section \ref{section:MBCE_MLSM} and using the decomposition
${\mb{X}}={\mb{L}}{\mb{S}}$, we summarize the resulting DD {\it CEEA-ML} detectors
for PSK and QAM based SM systems below.

\subsubsection{$M$-PSK Constellation}
For a PSK-SM MIMO system, the {\it CEEA-ML detector} is
\begin{IEEEeqnarray}{rCl}
\hat{\mb{X}}^{\textrm{ML}}_{dd}(k)=\arg
\underset{{s}_j\in\mathcal{A}_M,\:{\ell}_j
\in\mathbb{L}}{\widetilde\min}&&\log\det{\tilde{\mb{C}}_{psk}}\notag\\
&&+\mathcal{G}(\tilde{\mb{C}}_{psk},{\mb{y}}_s(k)-\tilde{\mb{m}}_{ssk})~~~
\label{eq:DDML_PSK}
\end{IEEEeqnarray}
where $\tilde{\mb{m}}_{ssk}=({\mb{L}}^{T}\otimes\mb{I}_{N_R})\mb{A}
\mathrm{vec}(\hat{\mb{H}})$,
${\tilde{\mb{C}}_{psk}}\defeq\frac{\sigma^{2}_{z}}{\varepsilon_s}\mb{I}_{BN_R}$ $+
({\mb{L}}^{T}\otimes\mb{I}_{N_R})\Big[\mb{I}_{N_RN_T}-\rho_{T}(1)\mb{A}\Big]$
$\mb{\Phi}({\mb{L}}^{\ast}\otimes\mb{I}_{N_R})$, and ${\mb{y}}_s(k)=
\mathrm{vec}\left(\mb{Y}(k){\mb{S}}^H\right)/\varepsilon_s$.

\subsubsection{$M$-QAM Constellation}
When $M$-QAM is used, (\ref{eq:ML_DD}) reduces to
\begin{IEEEeqnarray}{rCl}\hat{\mb{X}}^{\textrm{ML}}_{dd}(k)=\arg
\underset{{s}_j\in\mathcal{A}_M,\:{\ell}_j\in\mathbb{L}}{\widetilde\min}&&
N_R\log\det{\mb{E}}_{s}+\log\det{\tilde{\mb{C}}_{qam}}\notag\\
&&+\mathcal{G}(\tilde{\mb{C}}_{qam},{\mb{y}}_s(k)-\tilde{\mb{m}}_{ssk})~~~
\label{eq:DDML_QAM}
\end{IEEEeqnarray}
where $\tilde{\mb{C}}_{qam}\defeq$ $({\mb{L}}^{T}\otimes\mb{I}_{N_R})
\Big[\mb{I}_{N_RN_T}-\rho_{T}(1)\mb{A}\Big]
\mb{\Phi}({\mb{L}}^{\ast}\otimes\mb{I}_{N_R})$
$+\sigma^{2}_{z}({\mb{E}}_{s}^{-1}\otimes\mb{I}_{N_R})$, ${\mb{y}}_s(k)
\defeq \mathrm{vec}\left(\mb{Y}(k){\mb{S}}^H{\mb{E}}_s^{-1}\right)$, and ${\mb E}_s
\defeq {\mb{S}}{\mb{S}}^H$.

\begin{remark}
Compared with the MB-CE-aided detectors (cf. (\ref{eq:MBML_PSK}) and (\ref{eq:MBML_QAM})),
(\ref{eq:DDML_PSK}) and (\ref{eq:DDML_QAM}) require much less storage. The MB channel estimation
is performed every two frames and thus $2N-2$ estimated channel matrices have to be updated and
saved for subsequent signal detection. The corresponding $\mb{A}(k)$'s can be precalculated but
they have to be stored as well. The DD channel estimator aided detectors, on the other hand, need
to store two matrices, $\mb{A}$ and $\hat{\mb H}$, only. However, as shown in Section \ref{sec:sim},
the latter suffers from inferior performance. More detailed memory requirement comparison is provided
in Section \ref{sec:complexity}. $\square$
\end{remark}

\subsection{Two-Stage Detector for $M$-PSK SM Signals}
\label{subsec:CR_ML_DD}
We can reduce the search range of (\ref{eq:DDML_PSK}) from the $B$th Cartesian power of $\mathbb{L}
\times\mathcal{A}_M$ to $\mathbb{L}^B$ by adopting the two-stage approach of (\ref{eq:MB_2Stage})
that detects the antenna indices and then the transmitted symbols. The corresponding detector is
derived from maximizing the likelihood function
\begin{IEEEeqnarray}{rCl}
P\left[\mb{Y}(k)\Big|\mb{X}(k),\hat{\mb{H}}(k-1)\right]
\notag
\end{IEEEeqnarray}
with respect to $\mb{X}(k)$:
\begin{IEEEeqnarray}{rCl}\label{eq:SMML_PSK1}
&&\underset{{\mb{L}}\in\mathcal{L},\:
{\mb{s}}\in\mathcal{A}_M^B}{\max}\hspace{-.3em}
(\det \varepsilon_s{\tilde{\mb{C}}_{psk}})^{-\frac{1}{2}}
\exp\left[\frac{-\mathcal{G}(\tilde{\mb{C}}_{psk},\tilde{\mb{m}}_{ssk})}{2}\right.
\notag\\
&&\hspace{12.5em}\left.-\frac{{\mb{s}}^{T}{\mb{J}}(k){\mb{s}}^{*}}{2\varepsilon_s^{2}}
+\frac{\Re\{{\mb{b}}^H(k){\mb{s}}^{*}\}}{\varepsilon_s}\right]\notag\\
&&\approx\underset{{\mb{L}}\in\mathcal{L}}{\max}~
(\det \varepsilon_s{\tilde{\mb{C}}_{psk}})^{-\frac{1}{2}}
\exp\left[\frac{-\mathcal{G}(\tilde{\mb{C}}_{psk},\tilde{\mb{m}}_{ssk})}{2}\right]
\notag\\
&&\hspace{4.5em}\cdot\underset{\mb{s}\in\mathcal{A}_M^B}{\max}\:\exp\left[-\frac{1}{2\varepsilon_s^{2}}
{\mb{s}}^{T}{\mb{J}}(k){\mb{s}}^{*}
+\frac{1}{\varepsilon_s}\Re\{{\mb{b}}^H(k){\mb{s}}^{*}\}\right]\notag\\
&&=\underset{{\mb{L}}\in\mathcal{L}}{\max}\:
(\det \varepsilon_s{\tilde{\mb{C}}_{psk}})^{-\frac{1}{2}}
\exp\left[\frac{-\mathcal{G}(\tilde{\mb{C}}_{psk},\tilde{\mb{m}}_{ssk})}{2}\right.
\notag\\
&&\hspace{3em}\left.-\frac{\bar{\mb{s}}^T({\mb{L}}){\mb{J}}(k)
\bar{\mb{s}}^{*}({\mb{L}})}{2\varepsilon_s^{2}}+\frac{\Re\{{\mb{b}}^H(k)\bar{\mb{s}}^{*}({\mb{L}})
\}}{\varepsilon_s}\right],
\end{IEEEeqnarray}
where ${\mb{J}}(k)$, ${\mb{b}}(k)$, and $\bar{\mb{s}}({\mb{L}})$
are defined similarly to those in (\ref{eq:ML_2S_DD_J}), (\ref{eq:ML_2S_DD_b}) and
(\ref{eq:MB_NML_App}) except that now $\hat{\mb{H}}=\hat{\mb{H}}(k-1)$.
Replacing the likelihood function by its logarithm version, we obtain a
{\it two-stage detector} similar to (\ref{eq:MB_2Stage})
\begin{subequations}
\begin{IEEEeqnarray}{rCl}
\hat{\mb{L}}(k)&=&\arg\underset{{\mb{L}}\in\mathcal{L}}{\min}~
\log\det(\varepsilon_s{\tilde{\mb{C}}_{psk}})+\mathcal{G}(\tilde{\mb{C}}_{psk},\tilde{\mb{m}}_{ssk})\notag\\
&&\hspace{3.5em}+\frac{\bar{\mb{s}}^{T}({\mb{L}})
{\mb{J}}(k)\bar{\mb{s}}^{*}({\mb{L}})}{\varepsilon_s^{2}}
-\frac{2\Re\{{\mb{b}}^H(k)\bar{\mb{s}}^{*}
({\mb{L}})\}}{\varepsilon_s},\notag\\ \\
\hat{\mb{s}}(k)&=&\bar{\mb{s}}(\hat{\mb{L}}(k)).
\label{eq:DDSMML_redu}
\end{IEEEeqnarray}
\label{eq:DD_2Stage}
\end{subequations}

\section{ML Detection Without Spatial Correlation Information at Either Side}
\label{sec:AML}
The detector structures presented so far have assumed complete knowledge of the channel's spatial correlation
${\mb \Phi}$. As pointed out in \cite{ST_CorrCh}, when both sides of a link are richly scattered, the corresponding
spatial statistics can be assumed separable, yielding the Kronecker spatial channel
model \cite{KRCM}
\begin{IEEEeqnarray}{rCl}
    \mb{H}(k)=
     \mb{\Phi}_{R}^{\frac{1}{2}}\mb{H}_{w}(k)
     \mb{\Phi}_{T}^{\frac{1}{2}}
 \label{eq:ch_model}
\end{IEEEeqnarray}
with the spatial correlation matrix $\mb{\Phi}$ given by
\begin{IEEEeqnarray}{rCl}
\mb{\Phi}=
\mb{\Phi}_{T}\otimes \mb{\Phi}_{R},\label{eq:Kron_mod}
\end{IEEEeqnarray}
the Kronecker product of the spatial correlation matrix at the transmit side $\mb{\Phi}_{T}
=\mathbb{E}\{\mb{H}^T(k)\mb{H}^*(k)\}/\tr(\mb{\Phi}_T)$ and that at the receive side $\mb{\Phi}_{R}
=\mathbb{E}\{\mb{H}(k)\mb{H}^H(k)\}/\tr(\mb{\Phi}_R)$. When the latter is not available, we assume
that $\mb{\Phi}_{R}={\mb I}_{N_R}$ hence $\mb{H}(k)=\mb{H}_{w}(k)\mb{\Phi}_{T}^{\frac{1}{2}}$.
The assumption of uncorrelated receive antennas also implies
\begin{IEEEeqnarray}{rCl}
   P\Big[\mb{Y}(k)|\mb{X}(k),\hat{\mb{H}}(k)\Big]=\prod_{n=1}^{N_R}
P\Big[\underline{\mb{y}}_{n}(k)|\mb{X}(k),\hat{\underline{\mb{h}}}_{n}(k)\Big],~~~
\label{eq:like_approx}
\end{IEEEeqnarray}
where $\underline{\mb{y}}_{n}(k)$ is the sample (row) vector received by antenna $n$ at block $k$ and
$\hat{\underline{\mb{h}}}_{n}(k)$ the estimated channel vector between the $n$th receive antenna and
transmit antennas. As has been mention in Section I, we refer to a detector based on (\ref{eq:like_approx})
as the {\it ZRC detector}. An expression similar to (\ref{eq:like_approx}) for the case of
uncorrelated transmit antennas can be used to derive the ZTC detector.

\subsection{{MB-CE-Aided ZRC and ZTC Detectors}}
\label{section:ApproxMLMB}
\subsubsection{General ZRC/ZTC Detectors}
Define
\begin{IEEEeqnarray}{rCl}
\mb{z}_{1}^H&=&\underline{\mb{y}}_{n}(k)
=\underline{\mb{h}}_{n}(k)\mb{X}(k)+\underline{\mb{z}}_{n}(k),\notag\\
\mb{z}^{H}_{2}&=&\hat{\underline{\mb{h}}}_{n}(k)=\mb{t}^H(k)
\mb{T}^{-1}(k_{p})\times\notag\\
&&\:\left[\tilde{\mb{y}}_{n1}(k_p),\tilde{\mb{y}}_{n2}(k_p),
\cdots,\tilde{\mb{y}}_{nN_T}(k_p)\right]\notag
\end{IEEEeqnarray}
where $\mb Z(k)\defeq\left[\underline{\mb{z}}^T_1(k),\cdots,\underline{\mb{z}}^T_{N_R}(k)\right]^T$.
We immediately obtain the mean and covariance of $\underline{\mb{y}}_{n}(k)$ conditioned on
$\mb{X}(k)$ and $\hat{\underline{\mb{h}}}_{n}(k)$ as
\begin{subequations}
\begin{IEEEeqnarray}{rCl}
\bar{\mb{m}}_{n}^T(k)&=&\hat{\underline{\mb{h}}}_{n}(k)\mb{A}_{zrc}(k)\mb{X}(k),\label{eq:ApproxMLMBMean}\\
\bar{\mb{C}}_{zrc}(k)&=&\mb{\Sigma}_{11}-\mb{\Sigma}_{12}\mb{\Sigma}_{22}^{-1}\mb{\Sigma}_{12}^{H}
\notag\\
&=&\sigma^{2}_{z}\mb{I}_{N_T}\notag\\&&+\mb{X}^{H}(k)
\left[\mb{I}_{N_T}-\mb{A}(k)\left(\mb{t}^H(k)\mb{T}^{-1}(k_{p})\mb{q}(k)\right)^*
\right]\notag\\
&&\times\: \mb{\Phi}_T\mb{X}(k),\label{eq:ApproxMLMBCov}
\end{IEEEeqnarray}
\end{subequations}
where
$\mb{A}_{zrc}(k)=\mb{t}^H(k)\mb{T}^{-1}(k_{p})\mb{q}(k)\mb{\Phi}_T\mb{\Sigma}_{22}^{-1}$,
\begin{IEEEeqnarray}{rCl}
\mb{\Sigma}_{11}&=&\:\mb{X}^{H}(k)\mb{\Phi}_T\mb{X}(k)+\sigma^{2}_{z}\mb{I}_{N_T},\notag\\
\mb{\Sigma}_{12}
&=&\:\mb{t}^H(k)\mb{T}^{-1}(k_{p})\mb{q}(k)\mb{X}^{H}(k)\mb{\Phi}_T,\notag\\
\mb{\Sigma}_{22}&=&\:\nu(k)\mb{\Phi}_T+\sigma^{2}_{z}
\|\mb{t}^H(k)\mb{T}^{-1}(k_{p})\|_F^{2}\:\mb{I}_{N_T}\notag
\label{eq:ApproxMLMBSigma}
\end{IEEEeqnarray}
with
$\mb{q}(k)$
and $\nu(k)$ being defined in (\ref{eq:q_nu}).

The resulting {\it ZRC detector} is thus given by
\begin{align}
\hat{\mb{X}}^{\text{ZRC}}_{mb}(k)=&\arg\underset{{\mb{X}}\in\mathcal{X}}{\min}
N_{R}\log\det\bar{\mb{C}}_{zrc}(k)+\notag\\
&\tr\left\{\mathcal{G}\left(\bar{\mb{C}}_{zrc}(k),\left(\mb{Y}(k)-\bar{\mb{M}}_{zrc}(k)\right)^H\right)
\right\},
\label{eq:Approx_ST_MLMB}
\end{align}
where $\bar{\mb{M}}_{zrc}(k)=[\bar{\mb{m}}_{1}(k), \cdots, \bar{\mb{m}}_{n_R}(k)]^T
=\hat{\mb{H}}(k)\mb{A}_{zrc}(k)$ $\mb{X}(k)$.

It is verifiable that (\ref{eq:Approx_ST_MLMB}) can also be derived from substituting $\mb\Phi_R=\mb I_{N_R}$
into the {\it CEEA-ML} detection rule (\ref{eq:uni_ML_MB}) and applying (\ref{eq:Kron_mod}) with some additional
algebraic manipulations.

On the other hand, when the spatial correlation at the transmit side is not available,
we assume no {\it a priori} transmit spatial correlation, $\mb{\Phi}_T=\mb I_{N_T}$
whence $\mathbb{E}\{\mb{H}(k)\mb{H}^H(k)\}/\tr(\mb{\Phi}_R)=\mb{\Phi}_R$.
Substituting $\mb{\Phi}_T=\mb I_{N_T}$ into (\ref{eq:Kron_mod}) we obtain the
{\it ZTC} detector for a generic MB-CE-aided MIMO system with
\begin{IEEEeqnarray}{rCl}
\label{eq:MBMLMean_ZTC}
&&{\mb{m}}_{mb}(k)=\mb{t}^H(k)\mb{T}^{-1}(k_{p})\mb{q}(k)
\mathrm{vec}\left[\mb{\Phi}_R\left(\nu(k)\mb{\Phi}_R\right.\right.\notag\\
&&\left.\left.+\:\sigma^{2}_{z}\left\|\mb{t}^H(k)\mb{T}^{-1}(k_{p})\right\|_F^{2}
\mb{I}_{N_R}\right)^{-1}
\hat{\mb{H}}(k)\mb{X}(k)\right],
\end{IEEEeqnarray}
\begin{IEEEeqnarray}{rCl}
\mb{C}_{mb}(k)
&=&\sigma^{2}_{z}\mb{I}_{N_RN_T}
+\left[\mb{X}^{T}(k)\mb{X}^{\ast}(k)\right]\otimes\mb{\Phi}_R
\Big[\mb{I}_{N_R}\notag\\&&-\:|\mb{t}^H(k)\mb{T}^{-1}(k_{p})\mb{q}(k)|^2
\Big(\nu(k)\mb{I}_{N_R}\notag\\&&
+\:\sigma^{2}_{z}\left\|\mb{t}^H(k)\mb{T}^{-1}(k_{p})\right\|_F^{2}
\mb{\Phi}_R^{-1}\Big)^{-1}
\Big]\notag\\
&\approx&\mb{I}_{N_T}\otimes\bar{\mb{C}}_{ztc}(k),\label{eq:ztcapp}
\end{IEEEeqnarray}
where we use $\mb{X}(k)\mb{X}^{H}(k)\approx\varepsilon_s\mb{I}_{N_T}$ and
define
\begin{IEEEeqnarray}{rCl}
&&\bar{\mb{C}}_{ztc}(k)\defeq
\sigma^{2}_{z}\mb{I}_{N_R}
+\varepsilon_s\mb{\Phi}_R
\Big[\mb{I}_{N_R}-|\mb{t}^H(k)\mb{T}^{-1}(k_{p})\mb{q}(k)|^2\notag\\
&&\mb{\Phi}_R\Big(\sigma^{2}_{z}\left\|\mb{t}^H(k)\mb{T}^{-1}(k_{p})\right\|_F^{2}
\mb{I}_{N_R}+\nu(k)\mb{\Phi}_R\Big)^{-1}\Big].\label{eq:Cmb_ztc}
\end{IEEEeqnarray}
to obtain (\ref{eq:ztcapp}).
As a result, $\det\mb{C}_{ztc}(k)=\left(\det\bar{\mb{C}}_{ztc}(k)\right)^{N_T}$ and
\begin{IEEEeqnarray}{rCl}
&&\hspace{-2em}
\mathcal{G}(\mb{C}_{mb}(k),\mathrm{vec}(\mb{Y}(k))-{\mb{m}}_{mb}(k))
\notag\\
&=&\tr\left\{\mathcal{G}(\bar{\mb{C}}_{ztc}(k),\mb{Y}(k)-\bar{\mb{M}}_{ztc}(k))\right\},\notag
\end{IEEEeqnarray}
where
\begin{IEEEeqnarray}{rCl}
&&\bar{\mb{M}}_{ztc}(k)=\mb{t}^H(k)\mb{T}^{-1}(k_{p})\mb{q}(k)\cdot
\mb{\Phi}_R\Big(\nu(k)\mb{\Phi}_R\notag\\&&
+\:\sigma^{2}_{z}\left\|\mb{t}^H(k)\mb{T}^{-1}(k_{p})\right\|_F^{2}
\mb{I}_{N_R}\Big)^{-1}
\hat{\mb{H}}(k)\mb{X}(k).\label{eq:Mmb_ztc}
\end{IEEEeqnarray}
The resulting {\it ZTC detector} is given by
\begin{IEEEeqnarray}{rCl}
\hat{\mb{X}}^{\textrm{ZTC}}_{mb}(k)=\arg\underset{\mb{X}\in\mathcal{A}_{M^+}^{N_T\times B}}
{\min}&&\tr\left\{\mathcal{G}(\bar{\mb{C}}_{ztc}(k),\mb{Y}(k)-\bar{\mb{M}}_{ztc}(k))\right\}
\notag\\
&&+N_T\log\det {\bar{\mb{C}}_{ztc}}(k)
\label{eq:ZTC_MB}
\end{IEEEeqnarray}
which is of comparable computational complexity as that of the {\it ZRC detector}.

When both the transmit and receive correlations are unknown, the conditional covariance matrix
in the detection metric degenerates to an identity matrix scaled by a factor that is a function
of the noise variance, channel's temporal correlation and block index $k$.
Both {\it ZRC} and {\it ZTC} receivers can be further simplified by the two-stage approach,
we focus on deriving reduced complexity {\it ZRC detectors} only; the {\it ZTC} counterparts
can be similarly obtained.

\subsubsection{Two-Stage ZRC $M$-PSK Detector}
Two-stage ZRC and ZTC detectors can be obtained by following the derivation given in Section
\ref{section:MBCE_MLSM}. We present the two-stage ZRC detector in the following and omit the
corresponding ZTC detector. Using the decomposition, $\mb{X}=\mb{ L}\mb S$, we express the
{\it ZRC} receiver (\ref{eq:Approx_ST_MLMB}) for an $M$-PSK SM system as
\begin{IEEEeqnarray}{rCl}
\hat{\mb{X}}^{\text{ZRC}}_{mb}(k)&=&
\arg\underset{(\mb{s},\mb{L})\in\mathcal{A}_M^B\times\mathcal{L}}{\widetilde{\min}}
N_{R}\log\det(\varepsilon_s\bar{\mb{C}}_{ssk}(k))
\notag\\
&&+\tr\Big\{\mathcal{G}\Big(\bar{\mb{C}}_{ssk}(k),
\left(\mb{Y}_s(k)-\bar{\mb{M}}_{ssk}(k)\right)^H\Big)
\Big\}~~~~~
\label{eq:Approx_ST_MLMB_ssk}
\end{IEEEeqnarray}
where $\mb{Y}_s(k)={\mb{Y}(k){\mb{S}}^{H}}/{\varepsilon_s}$, $\bar{\mb{M}}_{ssk}(k)=\hat{\mb{H}}(k)
\mb{A}(k){\mb{L}}$, and
\begin{align}
&\bar{\mb{C}}_{ssk}(k)=\frac{\mb{\sigma}^{2}_{z}}{\varepsilon_s}\mb{I}_{N_T}+{\mb{L}}^{H}\times\notag\\
&\left[\mb{I}_{N_{T}}-\mb{A}(k)\mb{t}^H(k)\mb{T}^{-1}(k_{p})\mb{q}(k)
\right]\mb{\Phi}_T{\mb{L}}.\label{eq:Cmb_zrc2s}
\end{align}
Decision rule for separate antenna index and modulated symbol detection can be shown to be given by
\begin{subequations}
\begin{IEEEeqnarray}{rCl}
\hat{\mb{L}}(k)&=&\arg
\underset{{\mb{L}}\in\mathcal{L}}{\min}\:
{N_R}\log\det(\varepsilon_s
\bar{\mb{C}}_{ssk}(k))\notag\\
&&+\tr\left\{\mathcal{G}\left(\bar{\mb{C}}_{ssk}(k),
\bar{\mb{M}}_{ssk}(k)^{H}\right)\right\}\notag\\
&&+\frac{\bar{\mb{s}}^{H}({\mb{L}}){\mb{J}}(k)
\bar{\mb{s}}({\mb{L}})}{\varepsilon_s^{2}}-\frac{2\Re\{{\mb{b}}^T(k)\bar{\mb{s}}({\mb{L}})\}}{\varepsilon_s},\\
\hat{\mb{s}}(k)&=&\bar{\mb{s}}(\hat{\mb{L}}(k)),
\end{IEEEeqnarray}
\label{eq:MB_2SZRC}%
\end{subequations}
where the entries of ${\mb{b}}(k)$ are the diagonal terms of
$\mb{Y}^{H}(k)\bar{\mb{M}}_{ssk}\bar{\mb{C}}_{ssk}^{-1}(k)$, $\bar{\mb{s}}({\mb{L}})
=\mathcal{Q}_{\mathcal{A}_{M}}\left(\varepsilon_s({\mb{b}}^T(k){\mb{J}}^{-1}(k))^{H}\right)$,
and
${\mb{J}}(k)=\bar{\mb{C}}_{ssk}^{-1}(k)$ $\odot(\mb{Y}^{H}(k)\mb{Y}(k))^{*}$.

\subsection{DD-CE-Aided ZRC and ZTC Detectors}
\label{section:Approx_ST_MLDD}
\subsubsection{General ZRC/ZTC Detectors}
Based on the ZRC assumption and the fact that in a DD system $\mb{X}(k)$ is
detected with the DD CEs obtained at block $k-1$,
we follow the procedure presented in the previous subsection with $\mb{z}_{1}^H=\underline{\mb{y}}_{n}(k)$ and
$\mb{z}_{2}^H=\hat{\underline{\mb{h}}}_n\defeq\hat{\underline{\mb{h}}}_{n}(k-1)$ $=\underline{\mb{h}}_{n}(k-1)\mb{G}_{1}(k-1)
 +\underline{\mb{z}}_{n}(k-1)\mb{G}_{2}(k-1)$,
where $\mb{G}_{1}(k)\defeq\mb{X}(k)\hat{\mb{X}}^{\dag}(k)$ and
$\mb{G}_{2}(k)\defeq\mb{X}^{\dag}(k)\mb{G}_{1}(k)$,
to obtain the covariance matrices
\begin{subequations}
\begin{IEEEeqnarray}{rCl}\label{eq:Approx_ST_MLDD_Sig}
\mb{\Sigma}_{11}
           &=&\mb{X}^{H}(k)\mb{\Phi}_T\mb{X}(k)+\sigma^{2}_{z}\mb{I}_B,\\
\mb{\Sigma}_{12}
           &=&\rho_{T}(1)\mb{X}^{H}(k)\mb{\Phi}_T\mb{G}_{1}(k-1),\\
\mb{\Sigma}_{22}&=&\mb{G}_{1}^{H}(k-1)\left[\mb{\Phi}_T+\sigma^{2}_{z}
           \left(\mb{X}(k-1)\mb{X}^H(k-1)\right)^{-1}\right]\notag\\
           & &\times\: \mb{G}_{1}(k-1)\notag\\
           &\approx&\mb{G}_{1}^{H}(k-1)\left(\mb{\Phi}_T
            +\frac{\sigma^{2}_{z}}{\varepsilon_s}\mb{I}_B\right)\mb{G}_{1}(k-1).
\end{IEEEeqnarray}
\end{subequations}
It follows immediately that, given $\mb{X}(k)$ and $\hat{\underline{\mb{h}}}_{n}(k-1)$,
$\underline{\mb{y}}_{n}(k)$ has 
mean $\mb{z}_{2}^H\mb{\Sigma}_{22}^{-1}\mb{\Sigma}_{12}^H
           ={\rho_{T}(1)}\hat{\underline{\mb{h}}}_n
           (\mb{\Phi}_T+\sigma^{2}_{z}/\varepsilon_s\mb{I}_{B})^{-1}\mb{\Phi}_T\mb{X}(k)$
and covariance matrix
\begin{align}
\tilde{\mb{C}}_{zrc}\defeq&\mb{\Sigma}_{11}-
\mb{\Sigma}_{12}\mb{\Sigma}_{22}^{-1}\mb{\Sigma}_{12}^{H}\notag\\
=&{\sigma}^{2}_{z}\mb{I}_B+\mb{X}^{H}(k)\left[\mb{I}_{N_T}-\rho_{T}^2(1)\left(\mb{I}_{N_T}
       +\frac{\sigma^{2}_{z}}{\varepsilon_s}\mb{\Phi}_T^{-1}\right)^{-1}\right]\notag\\
&\times \mb{\Phi}_T \mb{X}(k)
\label{eq:Cdd_ZRC}
\end{align}
where $\rho_{T}(1)$ is a prediction term used to alleviate the error propagation effect.
The corresponding {\it ZRC detector} is then given by
\begin{align}
\hat{\mb{X}}^{\text{ZRC}}_{dd}(k)=&\arg
\underset{{\mb{X}}\in\mathcal{X}}{\min}\:
N_{R}\log\det{\tilde{\mb{C}}_{zrc}}\notag\\
&+\tr\left[\mathcal{G}\left(\tilde{\mb{C}}_{zrc},\mb{Y}^H(k)-\tilde{\mb{M}}_{zrc}^H\right)\right],
\label{eq:ZRC_DD}
\end{align}
where $\tilde{\mb{M}}_{zrc}=\rho_{T}(1)\hat{\mb{H}}
\left(\mb{I}_{N_T}+\sigma^{2}_{z}/{\varepsilon_s}\mb{\Phi}_T^{-1}\right)^{-1}{\mb{X}}(k)$.
This {\it ZRC detector} can also be derived from the {\it CEEA-ML detector} (\ref{eq:ML_DD})
by using the approximation $\mb{\Phi}_R=\mb{I}_{N_R}$ and with some algebraic manipulations.


To derive the {\it ZTC detector} for a DD-CE-aided MIMO system, we replace $\mb\Phi$
by $\mb I_{N_T}\otimes\mb\Phi_R$ in (\ref{eq:ML_DD_subs}) and invoke
\begin{subequations}
\begin{IEEEeqnarray}{rCl}
  \mb A&=&\rho_T(1)(\mb I_{N_T}\otimes\mb\Phi_R)(\mb I_{N_T}\otimes\mb\Phi_R+\sigma_z^2\mb I_{N_RN_T})^{-1}\notag\\
  &=&\rho_T(1)\:\mb I_{N_T}\otimes\left(\mb\Phi_R(\mb\Phi_R+\sigma_z^2\mb I_{N_R})^{-1}\right)
\end{IEEEeqnarray}
and $\mb{X}(k)\mb{X}^{H}(k)\approx\varepsilon_s\mb{I}_{N_T}$ to obtain
\begin{IEEEeqnarray}{rCl}
\mb m_{dd}
  &=&\mathrm{vec}\left(\rho_T(1)\left(\mb\Phi_R(\mb\Phi_R+\sigma_z^2\mb I_{N_R})^{-1}\right)\hat{\mb{H}}\mb X(k)\right)\notag\\
  &\defeq&\mathrm{vec}(\tilde{\mb M}_{ztc}),\label{eq:Mdd_ZTC}\\
\mb C_{dd}
  &\approx&\mb I_{N_T}\otimes\Big(\sigma_z^2\mb I_{N_R}\notag\\
  &&+\:
  \varepsilon_s\left(\mb\Phi_R-\rho_T^2(1)(\mb I_{N_R}+\sigma_z^2\mb\Phi_R^{-1})^{-1}\mb\Phi_R\right)\Big)\notag\\
  &\defeq&\mb I_{N_T}\otimes\tilde{\mb C}_{ztc}.\label{eq:Cdd_ZTC}
\end{IEEEeqnarray}
\end{subequations}
Thus, $\det\mb C_{dd}=(\det\mb I_{N_T})^{N_R}\cdot(\det\tilde{\mb C}_{ztc})^{N_T}$ and
\begin{IEEEeqnarray}{rCl}
\mathcal{G}(\mb{C}_{dd},\mathrm{vec}(\mb{Y}(k))-{\mb{m}}_{dd})
=\tr\left\{\mathcal{G}(\tilde{\mb{C}}_{ztc},\mb{Y}(k)-\tilde{\mb{M}}_{ztc})\right\}.\notag
\end{IEEEeqnarray}
and from (\ref{eq:ML_DD}) we have the {\it DD-ZTC detector}
\begin{IEEEeqnarray}{rCl}
\hat{\mb{X}}^{\textrm{ZTC}}_{dd}(k)=\arg\underset{\mb{X}\in\mathcal{A}_{M^+}^{N_T\times B}}
{\min}&&\tr\left\{\mathcal{G}(\tilde{\mb{C}}_{ztc},\mb{Y}(k)-\tilde{\mb{M}}_{ztc})\right\}
\notag\\
&&+N_T\log\det{\tilde{\mb{C}}_{ztc}}
\label{eq:ZTC_DD}
\end{IEEEeqnarray}
\subsubsection{Two-Stage ZRC $M$-PSK Detector}
Due to space consideration, we present the two-stage ZRC detector only. We first define
\begin{subequations}
\begin{IEEEeqnarray}{rCl}
\tilde{\mb{M}}_{ssk}&=&
\rho_{T}(1)\hat{\mb{H}}\left(\mb{I}_{N_T}+\frac{\sigma^{2}_{z}}{\varepsilon_s}\mb{\Phi}_T^{-1}\right)^{-1}{\mb{L}},
\label{eq:Mdd_ZRC2S}\\
\tilde{\mb{C}}_{ssk}&=&\frac{\mb{\sigma}^{2}_{z}}{\varepsilon_s}\mb{I}_B
+{\mb{L}}^{H}\left[\mb{I}_{N_T}-\rho_{T}^{2}(1)\left(\mb{I}_{N_T}
+\frac{\sigma^{2}_{z}}{\varepsilon_s}\mb{\Phi}_T^{-1}\right)^{-1}\right]\notag\\
& & \cdot\: \mb{\Phi}_T {\mb{L}}.\label{eq:Cdd_ZRC2S}
\end{IEEEeqnarray}
\end{subequations}
Using the above definitions and the decomposition ${\bf X}={\bf LS}$, we obtain an alternate expression for
(\ref{eq:Cdd_ZRC}) and then rewrite (\ref{eq:ZRC_DD}) as
\begin{IEEEeqnarray}{rCl}
\hat{\mb{X}}^{\text{ZRC}}_{dd}(k)&=&
\arg\hspace{-.5em}\underset{(\mb{s},\mb{L})\in\mathcal{A}_M^B\times\mathcal{L}}{\widetilde{\min}}
\hspace{-.5em}
N_{R}\log\det(\varepsilon_s\tilde{\mb{C}}_{ssk})\notag\\
&&+\tr\Big\{
\mathcal{G}\Big(\tilde{\mb{C}}_{ssk},
\left(\mb{Y}_{s}(k)
-\tilde{\mb{M}}_{ssk}\right)^{H}\Big)\Big\}.
\label{eq:Approx_ST_MLDD_ssk}
\end{IEEEeqnarray}
The corresponding {\it two-stage ZRC detector} can then be derived as
\begin{subequations}
\begin{IEEEeqnarray}{rCl}
\hat{\mb{L}}(k)=\arg
\underset{{\mb{L}}\in\mathcal{L}}{\min}\:
&&{N_R}\log\det(\varepsilon_s
\tilde{\mb{C}}_{ssk})\notag\\
&&+\tr\Big\{
\mathcal{G}\left(\tilde{\mb{C}}_{ssk},\tilde{\mb{M}}_{ssk}^{H}\right)\Big\}
\notag\\
&&+\frac{\bar{\mb{s}}^{H}({\mb{L}}){\mb{J}}(k)
\bar{\mb{s}}({\mb{L}})}{\varepsilon_s^{2}}-\frac{2\Re\{{\mb{b}}^T(k)\bar{\mb{s}}({\mb{L}})\}}{\varepsilon_s}
,~~~~~~\\
\hat{\mb{s}}(k)=\bar{\mb{s}}(\hat{\mb{L}}(k))
\end{IEEEeqnarray}
\label{eq:DD_2SZRC}%
\end{subequations}
where $\bar{\mb{s}}({\mb{L}})=\mathcal{Q}_{\mathcal{A}_{M}}\left(\varepsilon_s({\mb{b}}^T
(k){\mb{J}}^{-1}(k))^{H}\right) $,
${\mb{b}}(k)$ equals to the diagonal of
$\mb{Y}^{H}(k)\tilde{\mb{M}}_{ssk}\tilde{\mb{C}}_{ssk}^{-1}$,
and
${\mb{J}}(k)=\tilde{\mb{C}}_{ssk}^{-1}\odot(\mb{Y}^{H}(k)\mb{Y}(k))^{*}$.

As the dimension of $\mb{A}_{zrc}(k)$ for the {\it ZRC detectors}
(\ref{eq:Approx_ST_MLMB}) and (\ref{eq:Approx_ST_MLMB_ssk}),
$N_T\times N_T$, is much smaller than that for the {\it CEEA-ML detectors}, $N_RN_T\times
N_RN_T$, the former class of detectors needs far less memory space. The detail comparison
of memory requirement and complexity of all detectors can be found in the next section. 

\section{Performance Analysis of CEEA-ML and Related Detectors}
\label{sec:theo}
The bit error rate (BER) performance of the SM detectors depends on the channel estimation method used
and, because of the assumed frame structure, is a function of the data block location index $k$; see
Fig. \ref{fig:pilot}. For the MB systems, the performance is better when $\min \{k, (N-k) \text{mod }N\}$
is small while for the DD counterparts, the performance degrades with increasing $k$. Let $\BER(k)$ be
the BER of the $k$th block then the average BER is
\begin{IEEEeqnarray}{rCl}
\BER_{\text{MB}}&=&\frac{1}{2N-2}\sum_{k=1,k\neq N}^{2N-1}\BER_{\text{MB}}(k),\label{eq:theo_BER_MB}\\
\BER_{\text{DD}}&=&\frac{1}{N-1}\sum_{k=1}^{N-1}\BER_{\text{DD}}(k).
\label{eq:theo_BER_DD}
\end{IEEEeqnarray}
For simplicity, the subscript for the channel estimator used shall be omitted unless necessary.
It is straightforward to show that \cite{CE_eff,BERfading} for {\it CEEA-ML detectors}
\begin{IEEEeqnarray}{rCl}
 \BER(k)&\leq&\frac{1}{2^{mB}}\frac{1}{mB}
 \sum_{(\mb{s},\mb{L})\in\mathcal{A}_M^B\times\mathcal{L}}
 \sum_{\underset{\mathcal{A}_M^B\times\mathcal{L}
       \setminus\{(\mb{s},\mb{L})\}}{(\mb{s}',\mb{L}')\in}}d_H\left(\mb X,\mb X'\right)\notag\\
       &&\hspace{6em}\cdot\: P_k\left\{\mb X\rightarrow\mb X'\right\}~~~~~~
\label{eq:UB}
\end{IEEEeqnarray}
where $\mb X=\mb L\cdot\Diag(\mb s)$, $\mb X'=\mb L'\cdot\Diag(\mb s')$, $d_H(\mb X,\mb X')$ denotes the
Hamming distance between the information bits carried by $\mb X$ and by $\mb X'$, and $P_k\{\mb X\rightarrow
\mb X'\}$ the averaged pairwise error probability (PEP) of detecting the transmitted signal $\mb X$ as
$\mb X'$. We derive the conditional PEP of the MB-CE-aided {\it CEEA-ML detectors} in the followings; that
for DD-CE-aided detectors can be similarly obtained.
\begin{IEEEeqnarray}{rCl}
&&\hspace{0em}P_k\left\{\mb X\rightarrow
\mb X'|\hat{\mb H}(k)\right\}\notag\\
&&=P_k\Big\{\text{\small{$\mathcal{G}(\mb{C}'_{mb}(k),
\mathrm{vec}(\mb{Y}(k))-\mb{m}'_{mb}(k))
+\log\det{\mb{C}'_{mb}}(k)$}}\notag\\
&&\hspace{1em}\text{\small{$<\mathcal{G}(\mb{C}_{mb}(k),
\mathrm{vec}(\mb{Y}(k))-{\mb{m}}_{mb}(k))
+\log\det{{\mb{C}}_{mb}}(k)|\hat{\mb H}(k)$}}\Big\}\notag\\
&&=P_k\Big\{(\mb y-\mb d)^H\mb D(\mb y-\mb d)-\bm\epsilon^H\mb D^{-1}\bm\epsilon\notag\\
&&\hspace{2.5em}+\:\mathcal{G}\left(\mb C'_{mb}(k),\mb m'_{mb}(k)\right)-\mathcal{G}\left(\mb C_{mb}(k),
\mb m_{mb}(k)\right)\notag\\
&&\hspace{2.5em}+\log\det\left(\mb{C}'_{mb}(k)\mb C_{mb}^{-1}(k)\right)<0|\hat{\mb H}(k)\Big\}
\label{eq:CompSq}\\
&&\defeq P_k\Big\{(\mb y-\mb d)^H\mb D(\mb y-\mb d)<\eta(\mb X,\mb X',k)|\hat{\mb H}(k)\Big\},\notag
\end{IEEEeqnarray}
where $\mb m'_{mb}(k)$ and $\mb C'_{mb}(k)$ are respectively obtained from (\ref{eq:m_mb}) and (\ref{eq:C_mb})
with $\mb X$ replaced by $\mb X'$ and (\ref{eq:CompSq}) is obtained by completing the square with $\mb D =
(\mb{C}'_{mb}(k))^{-1}-{\mb{C}}_{mb}^{-1}(k)$, $\mb y=\VEC(\mb Y(k))$, $\bm\epsilon=(\mb{C}'_{mb}(k))^{-1}
\mb m'_{mb}(k)-{\mb{C}}_{mb}^{-1}(k){\mb m}_{mb}(k)$, and $\mb d=\mb D^{-1}\bm\epsilon$. Based on (\ref{eq:MBML_Y}),
we have
\begin{IEEEeqnarray}{rCl}
  \mb y
  \sim\mathcal{CN}\left(\VEC\left(\hat{\mb H}(k)\mb X\right),\tilde{\mb{\Psi}}\right),
\end{IEEEeqnarray}
where $\tilde{\mb{\Psi}}\defeq\sigma^{2}_{z}\mb{I}_{N_RN_T}
+(\mb{X}^{T}\otimes\mb{I}_{N_R})\mb{\Psi}_{E}(k)(\mb{X}^*\otimes\mb{I}_{N_R})$.
Representing the Gaussian random vector $\mb y$ by $\tilde{\mb{\Psi}}^{\frac{1}{2}}\tilde{\mb y}+\VEC(\hat{\mb H}(k)\mb X)$
where $\tilde{\mb y}\sim\mathcal{CN}(\mb0_{N_RN_T},\mb I_{N_RN_T})$, we obtain an alternate quadratic form
\begin{IEEEeqnarray}{rCl}
  &&(\mb y-\mb d)^H\mb D(\mb y-\mb d)\notag\\
  &&=\text{\small{$\left(\tilde{\mb{\Psi}}^{\frac{1}{2}}\tilde{\mb y}
    +\VEC(\hat{\mb H}(k)\mb X)-\mb d\right)^H\mb D
    \left(\tilde{\mb{\Psi}}^{\frac{1}{2}}\tilde{\mb y}
    +\VEC(\hat{\mb H}(k)\mb X)-\mb d\right)$}}\notag\\
  &&=\left(\tilde{\mb y}
    +\tilde{\mb{\Psi}}^{-\frac{1}{2}}\left(\VEC(\hat{\mb H}(k)\mb X)-\mb d\right)\right)^H\notag\\
    &&\hspace{1.2em}\cdot\:\tilde{\mb{\Psi}}^{\frac{1}{2}}\mb D\tilde{\mb{\Psi}}^{\frac{1}{2}}
    \left(\tilde{\mb y}
    +\tilde{\mb{\Psi}}^{-\frac{1}{2}}\left(\VEC(\hat{\mb H}(k)\mb X)-\mb d\right)\right)\notag\\
  &&=\left(\mb U^H\tilde{\mb y}
    +\mb U^H\tilde{\mb{\Psi}}^{-\frac{1}{2}}\left(\VEC(\hat{\mb H}(k)\mb X)-\mb d\right)\right)^H\notag\\
    &&\hspace{1.2em}\cdot\:\mb\Lambda
    \left(\mb U^H\tilde{\mb y}
    +\mb U^H\tilde{\mb{\Psi}}^{-\frac{1}{2}}\left(\VEC(\hat{\mb H}(k)\mb X)-\mb d\right)\right)
    \defeq q_{nc},\notag
\end{IEEEeqnarray}
where $\mb U\mb\Lambda\mb U^H$ is the eigenvalue decomposition of $\tilde{\mb{\Psi}}^{\frac{1}{2}}
\mb D\tilde{\mb{\Psi}}^{\frac{1}{2}}$ with orthonormal matrix $\mb U$ and diagonal matrix $\mb\Lambda$
containing respectively the eigenvectors and corresponding eigenvalues. Since
$\mb U^H\tilde{\mb y}$ and $\tilde{\mb y}$ have the same distribution,
$q_{nc}$ is a noncentral quadratic form in Gaussian random variables and the CDF
\begin{equation}
P_k\Big\{q_{nc}<\eta(\mb X,\mb X',k)|\hat{\mb H}(k)\Big\}
\label{eq:cond_CDF}
\end{equation}
can be evaluated by the method proposed in \cite{QG} or the series expansion approach of \cite[Ch. 29]{StatBook}.

The average PEP is thus given by
\begin{IEEEeqnarray}{rCl}
&&P_k\left\{\mb X\rightarrow
\hat{\mb X}\right\}\label{eq:PWE}\\
&&=\int P_k\Big\{q_{nc}<\eta(\mb X,\mb X',k)|\hat{\mb H}(k)\Big\}P(\hat{\mb H}(k))\:\mathrm{d}\hat{\mb H}\notag\\
&&=\int P_k\Big\{q_{nc}<\eta(\mb X,\mb X',k)|\hat{\mb H}(k)\Big\}
\frac{e^{-\mathcal{G}(\tilde{\mb{\Phi}}(k),\VEC(\hat{\mb H}(k)))}}{\pi^{N_RN_T}\det(\tilde{\mb{\Phi}}(k))}
\mathrm{d}\hat{\mb H}\notag
\end{IEEEeqnarray}
where $\tilde{\mb{\Phi}}(k)\defeq\mathrm{E}\{\VEC(\hat{\mb H}(k))\VEC^H(\hat{\mb H}(k))\}$ is equal to (\ref{eq:MBMLSigma22}).
As the both sides of the inequality in (\ref{eq:cond_CDF}) depend on $\hat{\mb H}(k)$, the average over $\hat{\mb H}(k)$
(\ref{eq:PWE}) can only be computed by numerical integration. The BER performance of the {\it ZRC}, {\it ZTC} and general
MIMO {\it CEEA-ML detectors} can also be analyzed by the same approach with the corresponding PEP derived from different
spatial correlation structures. 

\section{Complexity and Memory Requirement of Various SM Detectors}
\label{sec:complexity}
We now compare the computational complexity and memory requirement of the detectors derived so far.
The memory space is used to store the required items involved in the
detection metrics and is divided into two categories: i) fixed and ii) dynamic. The former stores
the items that are independent of the received samples and/or updated channel estimates and can be
calculated offline. The latter specifies those vary with the received samples. Take the {\it ML detector}
(\ref{eq:uni_ML_MB}) for example, to achieve fast real-time detection, we pre-calculate and store
$\mb C^{-1}_{mb}(k)$, $\det\mb C_{mb}(k)$ and $\mb m_{mb}(k)$ for all candidate signals ($\mb{X}$)
and all $k$ in two consecutive frames. These items are time-invariant, independent of $\mb Y(k)$
or $\hat{\mb H}(k)$. The dynamic part refers to the received samples (in two frames) which have to
be buffered before being used to compute the MB channel estimate and detecting the associated signals.
As $(\mb{X}^{T}(k)\otimes\mb{I}_{N_R})\mb{A}(k)$ for all candidate $\mb{X}$ can be precalculated
and stored, the complexity to compute (\ref{eq:m_mb}) is only $\mathcal{O}(M^{N_TB} N_R^2 N_T^2)$
complex multiplications per block. The remaining complexity is that of computing $\mathcal{G}(\mb{C}_{mb}(k),
\mathrm{vec}(\mb{Y}(k))-{\mb{m}}_{mb}(k))$.

As the {\it CEEA-ML detector} degenerates to the {\it ZRC detector} by setting $\mb\Phi_R=\mb I_{N_R}$,
the dimensions of the correlation-related terms can be reduced by a factor of $N_R$. This reduction directly
affects both computing complexity and memory requirement. The complexity of the {\it two-stage detector}
is only $1/M^B$ of its single-stage counterpart because the parallel search on both transmit antenna index
and data symbol has been serialized. However, the memory required remains unchanged.

For the {\it mismatched detectors} (\ref{eq:MM_D}), besides the dynamic memory to store $\hat{\mb H}$
and $\mb Y(k)$, the fixed memory, which consists mainly of those for storing the terms the channel estimators
need, is relative small and usually dominated by the memory to store the statistics for the {\it CEEA-ML detectors}.
A minimal complex multiplication complexity of $\mathcal{O}(M^{N_T}B N_R N_T)$ is called for, since without
channel statistics data of each channel use can be detected separately, i.e., $\hat{\mb{X}}^\text{MM}(k)\defeq
[\hat{\mb{x}}_1^\text{MM}(k),\cdots,\hat{\mb{x}}_B^\text{MM}(k)]$ with \[\hat{\mb{x}}_j^\text{MM}(k)=\arg
\underset{\mb{x}\in\mathcal{A}_{M^+}^{N_T}} {\min}\|\mb{y}_j(k)-\hat{\mb{H}}\mb{x}\|_F^2.\]

The detectors using DD channel estimates needs significantly less memory than those using MB ones as
they do not jointly estimate the channel of several blocks and (\ref{eq:ML_DD}) indicates that
$\mb C_{dd}^{-1}$ is independent of the block index. We summarize the computing complexities and
the required memory spaces of various detectors in Tables \ref{table} and \ref{MEMtable}.

%

\renewcommand{\arraystretch}{1.3}
\begin{table}[t]
\caption{Numbers of complex multiplications involved to detect a data block for various detectors and modulation schemes where $\gamma_1=M^{N_T}$ and $\gamma_2=M^{B}$}
 \centering
 \tabcolsep 0.01in
 \begin{tabular}{|c|c|c|c|}
  \hline
   & General MIMO & SM & Two-stage \\ \hline
  Mismatched & $\mathcal{O}(\gamma_1 B N_R N_T)$ & $\mathcal{O}(M B N_R N_T)$ & N/A\\ \hline
  CEEA-ML & $\mathcal{O}(\gamma_1^{B} N_R^2 N_T^{2})$ & $\mathcal{O}(\gamma_2 N_T^{B+2} N_R^2)$ & $\mathcal{O}(N_T^{B+2} N_R^2)$\\ \hline
  ZRC & $\mathcal{O}(\gamma_1^{B}N_R N_T^2)$ & $\mathcal{O}(\gamma_2 N_T^{B+2}N_R )$ & $\mathcal{O}(N_T^{B+2}N_R)$ \\ \hline
 \end{tabular}\label{table}
\end{table}
\renewcommand{\arraystretch}{1}

We conclude that, among all the proposed detectors, the {\it ZRC} (or {\it ZTC}) {\it detectors} are the
most desirable as they require the minimal computation and memory to achieve satisfactory
detection performance. Although the {\it mismatched detector} is the least complex and requires
only comparable memory as {\it ZRC} (or {\it ZTC}) {\it detectors} do, its performance, as shown in the following
section, is much worse than that of the proposed detectors in some cases.

\renewcommand{\arraystretch}{1.3}
\begin{table}[t]
\caption{Size of memory (in numbers of complex values) of various detectors to store required items to detect a data block where $\gamma=M^{N_TB}$, $N_T^B$, and $M^{B}N_T^B$ for general MIMO, PSK-SM, and QAM-SM signals, respectively}
 \centering
 \tabcolsep 0.02in
 \begin{tabular}{|C{.6in}C{.55in}|C{.9in}|C{.9in}|}
  \hline %
  & & MB channel estimates & DD channel estimates \\ \hline
  \multicolumn{1}{|c|}{\multirow{2}{*}{Mismatched}} & Fixed & 
  $\mathcal{O}(N)$ & $\mathcal{O}(B)$\\ \cline{2-4}
  \multicolumn{1}{|c|}{} & Dynamic & $\mathcal{O}(NN_RN_T)$ & $\mathcal{O}(N_RN_T)$ \\ \hline
  \multicolumn{1}{|c|}{\multirow{2}{*}{CEEA-ML}} & Fixed & $\mathcal{O}(\gamma N N_R^2 N_T^2)$ & $\mathcal{O}(\gamma N_R^2 N_T^2)$  \\ \cline{2-4}
  \multicolumn{1}{|c|}{} & Dynamic & $\mathcal{O}(N N_R N_T)$ & $\mathcal{O}(N_R N_T)$ \\ \hline
  \multicolumn{1}{|c|}{\multirow{2}{*}{ZRC}} & Fixed & $\mathcal{O}(\gamma N N_T^2)$ & $\mathcal{O}(\gamma N_T^2)$  \\ \cline{2-4}
  \multicolumn{1}{|c|}{} & Dynamic & $\mathcal{O}(N N_R N_T)$ & $\mathcal{O}(N_R N_T)$ \\ \hline
 \end{tabular}
 \label{MEMtable}
\end{table}
\renewcommand{\arraystretch}{1}

\section{Simulation Results}\label{sec:sim}
In this section, the BER performance of the detectors we derived is studied through computer simulations.
We use the S-T channel model \cite{ST_CorrCh} and assume that uniform linear
arrays (ULAs) are deployed on both sides of the link. The spatial correlation follows the Kronecker
model (\ref{eq:ch_model}) so that (\ref{non_Kron_rhoS}) can be written as
$\rho_{S}(i-m,j-n)={[\mb{\Phi}_T]_{jn}}\cdot
{[\mb{\Phi}_R]_{im}}=\rho_{S}(0,j-n)\cdot\rho_{S}(i-m,0)$.
When the angle-of-arrivals (AoAs) and angle-of-departures (AoDs) are uniformly distributed in $(0,2\pi]$,
we have \cite{ST_CorrCh} 
\begin{equation}
  \rho_{S}(\ell-\ell',0)=\rho_{S}(0,\ell-\ell')=J_{0}(2\pi(\ell-\ell')\delta/\lambda),
  \label{eq:SpCorr_Bessel}
\end{equation}
where $J_0(\cdot)$ is the zeroth-order Bessel function of the first kind, $\delta$ the
antenna spacing for both transmitter and receiver and $\lambda$ the signal wavelength.
On the other hand, if the AoAs and AoDs have limited angle spreads \cite{Gauss_AoA}, the
spatial correlation can be expressed as
\begin{equation}
[\mb\Phi_R]_{ij}=r^{|i-j|},~~[\mb\Phi_T]_{ij}=t^{|i-j|}
\label{eq:SpCorr_Exp}
\end{equation}
with $0\leq r,t<1$. The above exponential model has been widely applied for MIMO system evaluation
\cite{METIS} and proven to be consistent with field measurements \cite{ExpCorr}.  We use both
(\ref{eq:SpCorr_Exp}) and the isotropic model (\ref{eq:SpCorr_Bessel}). As for time selectivity,
we assume it is characterized by the well-known Jakes' model \cite{jakemodel}
 \begin{IEEEeqnarray}{rCl}
    \rho_{T}(k-\ell)
    &=&J_{0}(2\pi f_{D}(k-\ell)B T_{s}), \label{T_corr}
\end{IEEEeqnarray}
where $f_D$ is the maximum Doppler frequency and $T_s$ the symbol duration. The frame structure
we consider is shown in Fig. \ref{fig:pilot} where a pilot block of the form ${\mb I}_{N_T}$
(hence $B=N_T$) is placed at the beginning of each frame. The resulting pilot density is very low
when compared with that of the cell-specific reference (CRS) signal in the 3GPP LTE specifications
\cite{LTE}. If we follow the symbol duration defined in \cite{LTE}, the pilot densities are only
$1/6$ and $1/3$ of that used in 3GPP LTE. In the remainder of this section, we consider SM and SMX
MIMO systems of different system parameters, including numbers of antennas $N_R$ and $N_T$,
constellation size $M$, and frame size $N$, resulting in various transmission rates. If the pilot
overhead is taken into account, the effective rate will be $(N-1)/N$ of the nominal value. Note
that we consider only uncoded systems, its performance can easily be improved by two or three
orders of magnitude using a proper forward error correcting code.


%

\begin{figure}
  \centering
  \mbox{\subfloat[]{\label{subfig:MB_4x4_M4N5N10_lamb05} \includegraphics[width=3.6in]{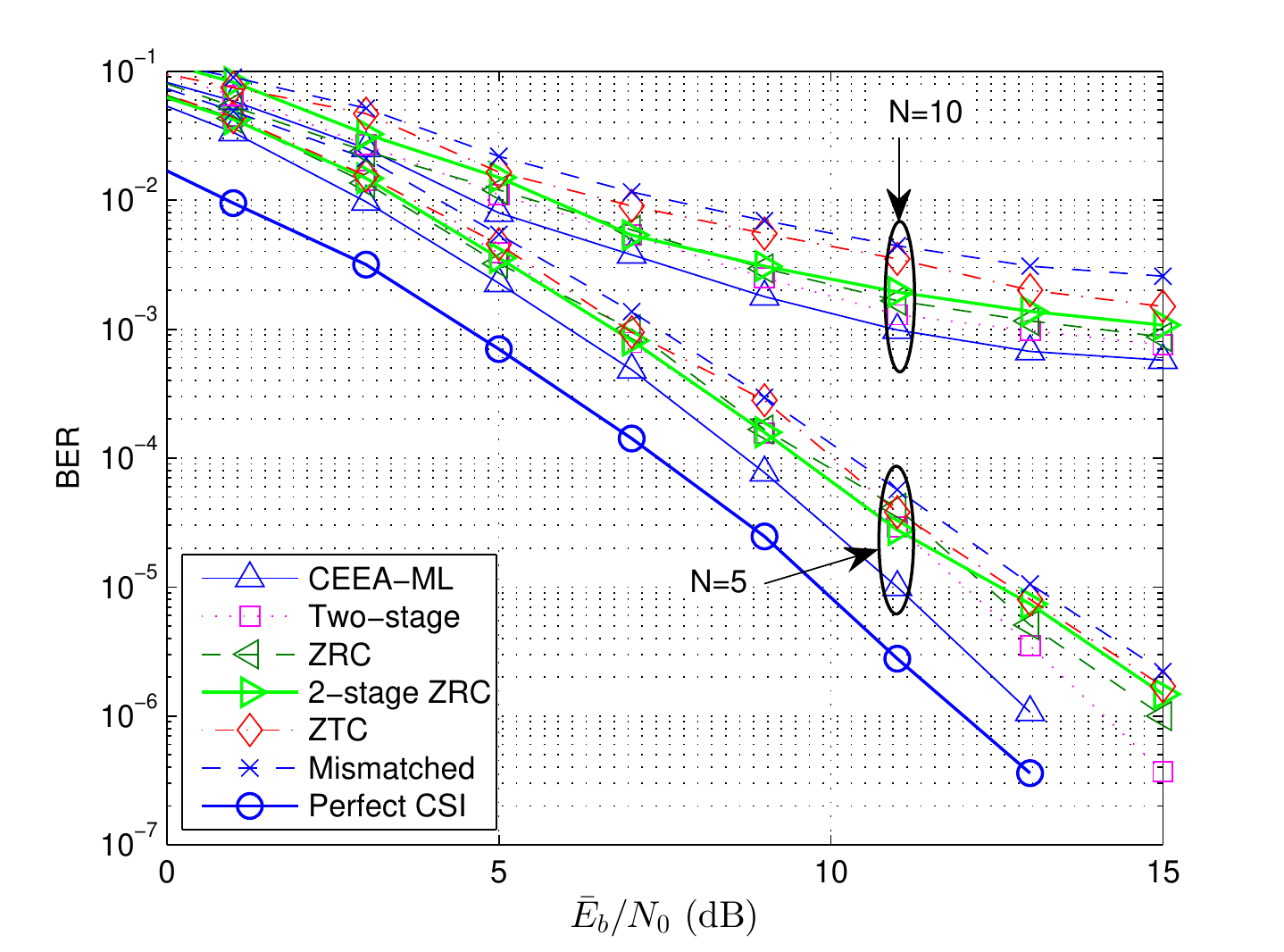}}}
  \mbox{\subfloat[]{\label{subfig:MB_4x4_M4N5N10_lamb1} \includegraphics[width=3.6in]{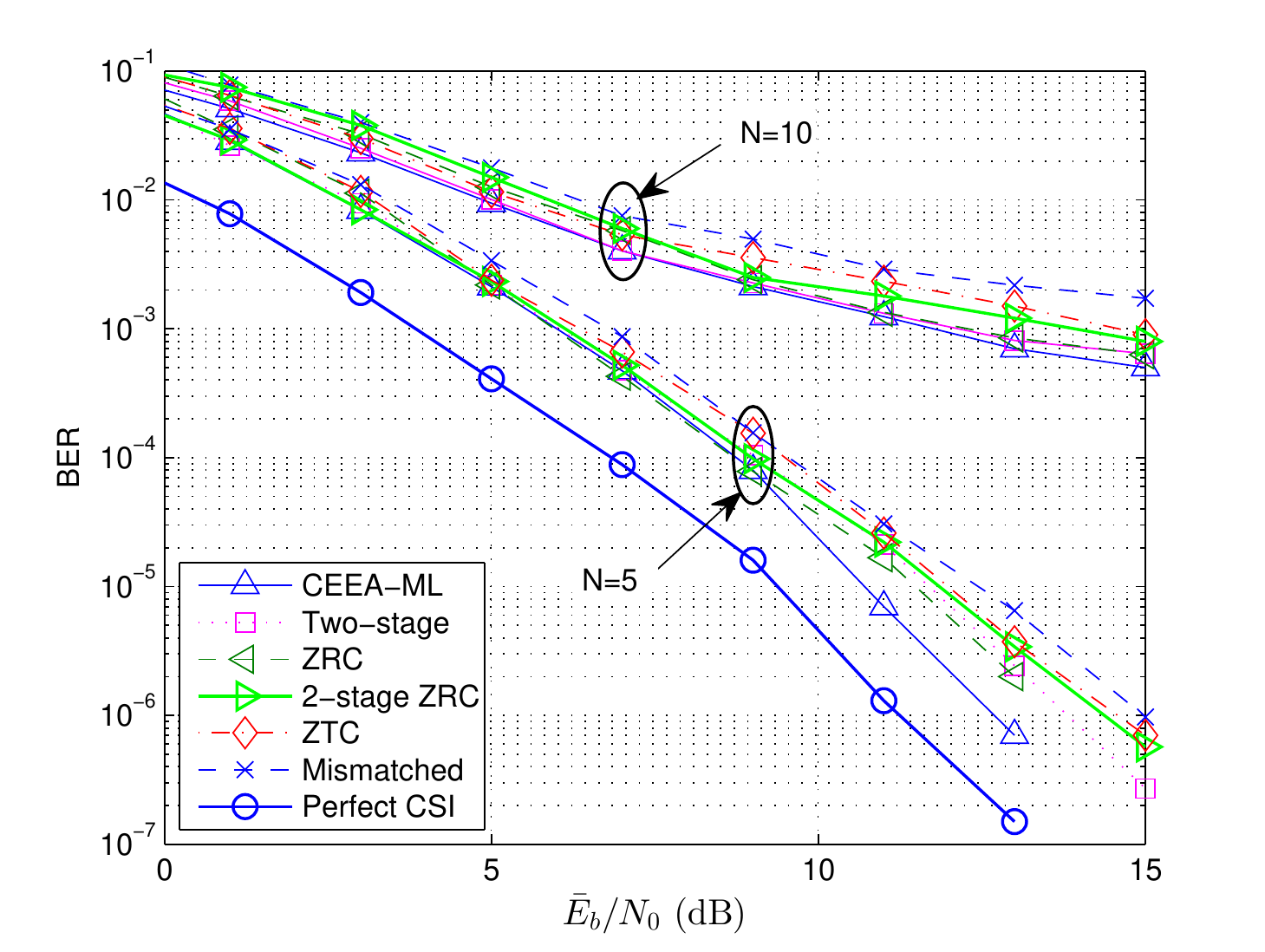}}}
        \caption{Impact of the frame size and antenna spacing on the performance of
        MB-CE-aided SM detectors in a Bessel correlated channel ($f_DT_s=0.01$):
        (a) $\delta=0.5\lambda$ (b) $\delta=1\lambda$; $M=4$, $N_T=N_R=B=4$.}
         \label{fig:MB_4x4_M4N5N10_Bes}
\end{figure}

%

\begin{figure}
  \centering
  \mbox{\subfloat[]{\label{subfig:MB_4x4_M4N5N10_r08} \includegraphics[width=3.6in]{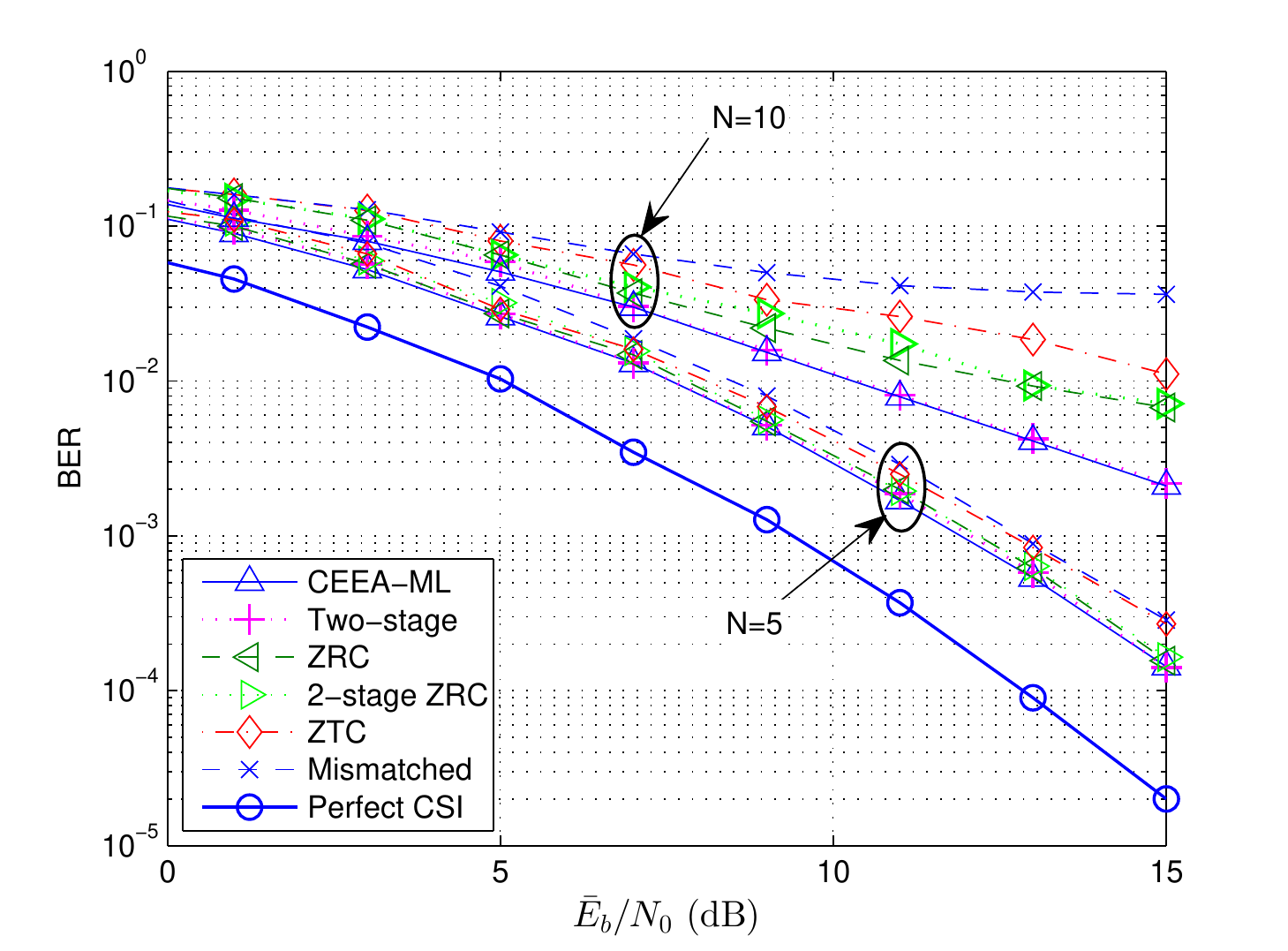}}}
  \mbox{\subfloat[]{\label{subfig:MB_4x4_M4N5N10_r05} \includegraphics[width=3.6in]{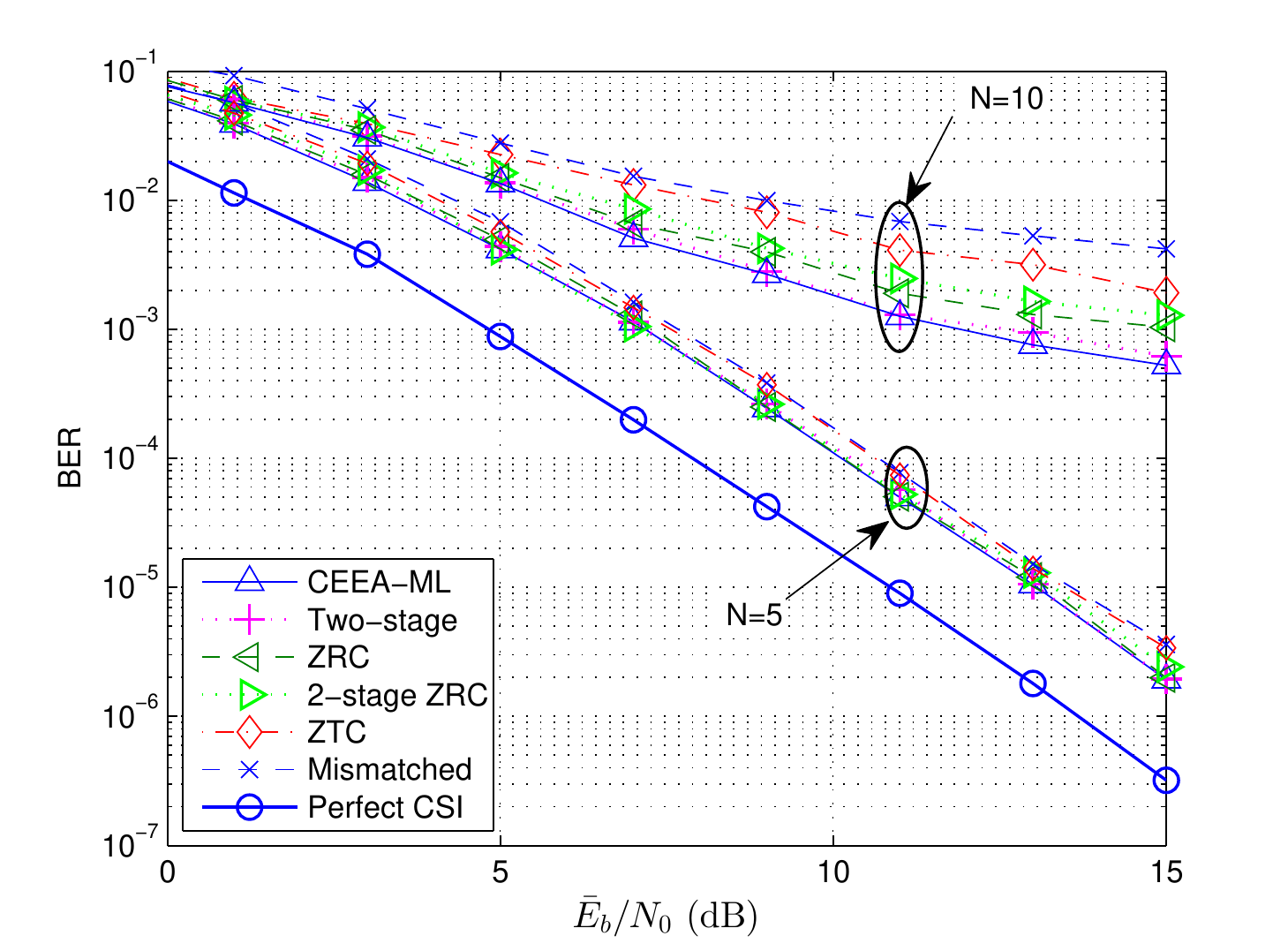}}}
        \caption{Impacts of frame size and spatial correlation on the performance of
        MB-CE-aided SM detectors in exponentially-correlated channels: (a) $r=t=0.8$;
        (b) $r=t=0.5$; $M=4$, $N_T=N_R=B=4$, $f_DT_s=0.01$.}
         \label{fig:MB_4x4_M4N5N10_Exp}
\end{figure}

\subsection{MB-CE-Aided SM Detectors}
The performance of MB-CE-aided CEEA detectors is presented in Figs. \ref{fig:MB_4x4_M4N5N10_Bes}
and \ref{fig:MB_4x4_M4N5N10_Exp} where we plot the BER performance of the {\it CEEA-ML} (\ref{eq:MBML_PSK}),
{\it mismatched}, and suboptimal detectors as a function of $\bar{E}_b/N_0$, $\bar{E}_b$ being
the average received bit energy per antenna. The suboptimal detectors include the {\it two-stage}
(\ref{eq:MB_2Stage}), {\it ZRC} (\ref{eq:Approx_ST_MLMB}), {\it ZTC} (\ref{eq:ZTC_MB}), and the
{\it two-stage ZRC detectors} (\ref{eq:MB_2SZRC}). While the ML and suboptimal detectors outperform the
{\it mismatched} one, the {\it ZTC detector} suffers slightly more performance degradation with respect to
(w.r.t.) the ML detector than the {\it ZRC} one does for it is obtained by using the extra approximation
$\mb{X}(k)\mb{X}^{H}(k)\approx \varepsilon_s\mb{I}_{N_T}$. The effect of spatial correlation can be found
by comparing the curves corresponding to $N=10$ (frame duration $40T_s$). When the spatial correlation
follows (\ref{eq:SpCorr_Bessel}) with $\delta=0.5$ or $1\lambda$, the correlation value is relatively low
and the knowledge of this information gives limited performance gain. But if the correlation is described
by (\ref{eq:SpCorr_Exp}) with $r=t=0.8$ or $0.5$, the {\it CEEA-ML} and its low-complexity variations
outperform the {\it mismatched detector} significantly. With $N=10$ the {\it two-stage detector}, which
requires a much lower complexity, suffers only negligible degradation w.r.t. its {\it CEEL-ML} counterpart.

Note that as the spatial correlation increases, it becomes more difficult for an SM detector to resolve
spatial channels (different $\mb h_j$'s) and thus the detection performance degrades accordingly. This holds for detectors
with perfect CSIR and those using the MB or DD CEs. Neglecting CSI error and channel correlation cause more
performance loss for channels with stronger correlations as can be found by comparing the mismatch losses.
Higher spatial correlation also causes larger performance degradation for the {\it ZRC} and {\it ZTC
detectors} which lack one side's spatial information. The effect of a shorter frame ($N=5$) can be found in
the same figures as well. As the CSI error is reduced, the performance gain, which is proportional to the
CSI error, becomes less impressive.

The perfect CSIR ML detector (\ref{eq:optMis}) does not need the channel's spatial and/or time correlation
information and whose performance is insensitive to time selectivity. For other detectors, the CSI error increases
with a larger $f_DT_s$ and/or a sparser pilot density and so is their performance degradations. For example, from
(\ref{T_corr}) we find that, for $f_DT_s=0.01$, the $50\%$-coherence time is approximately $24.2 T_s$ and thus
with the frame size $N=10$, each antenna receives a pilot symbol every other $39 T_s$ which is too sparse to
track the channel's temporal variation. Although knowing the resulting CSI error statistics does help reducing the
performance loss, increasing the pilot density to reduce the CSI error is much more efficient.
Increasing the pilot density by two-fold ($N=5$) recovers most losses w.r.t. the CEEA-ML detector for both channels.

%
%

\begin{figure}
  \centering
  \mbox{\subfloat[]{\label{subfig:MB_CorrEst_4x4_M4N5N10_r08} \includegraphics[width=3.6in]{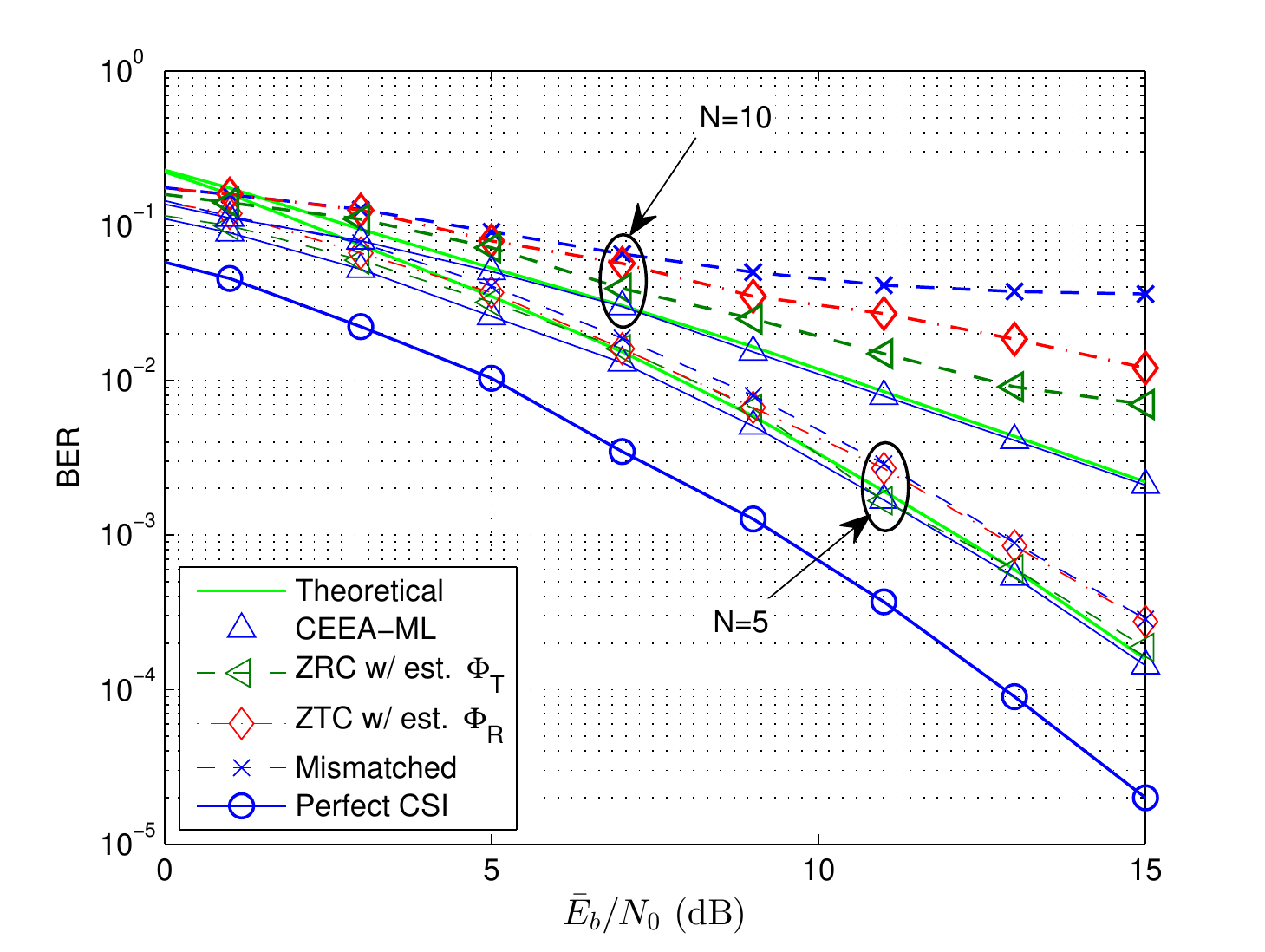}}}
  \mbox{\subfloat[]{\label{subfig:MB_CorrEst_4x4_M4N5N10_r05} \includegraphics[width=3.6in]{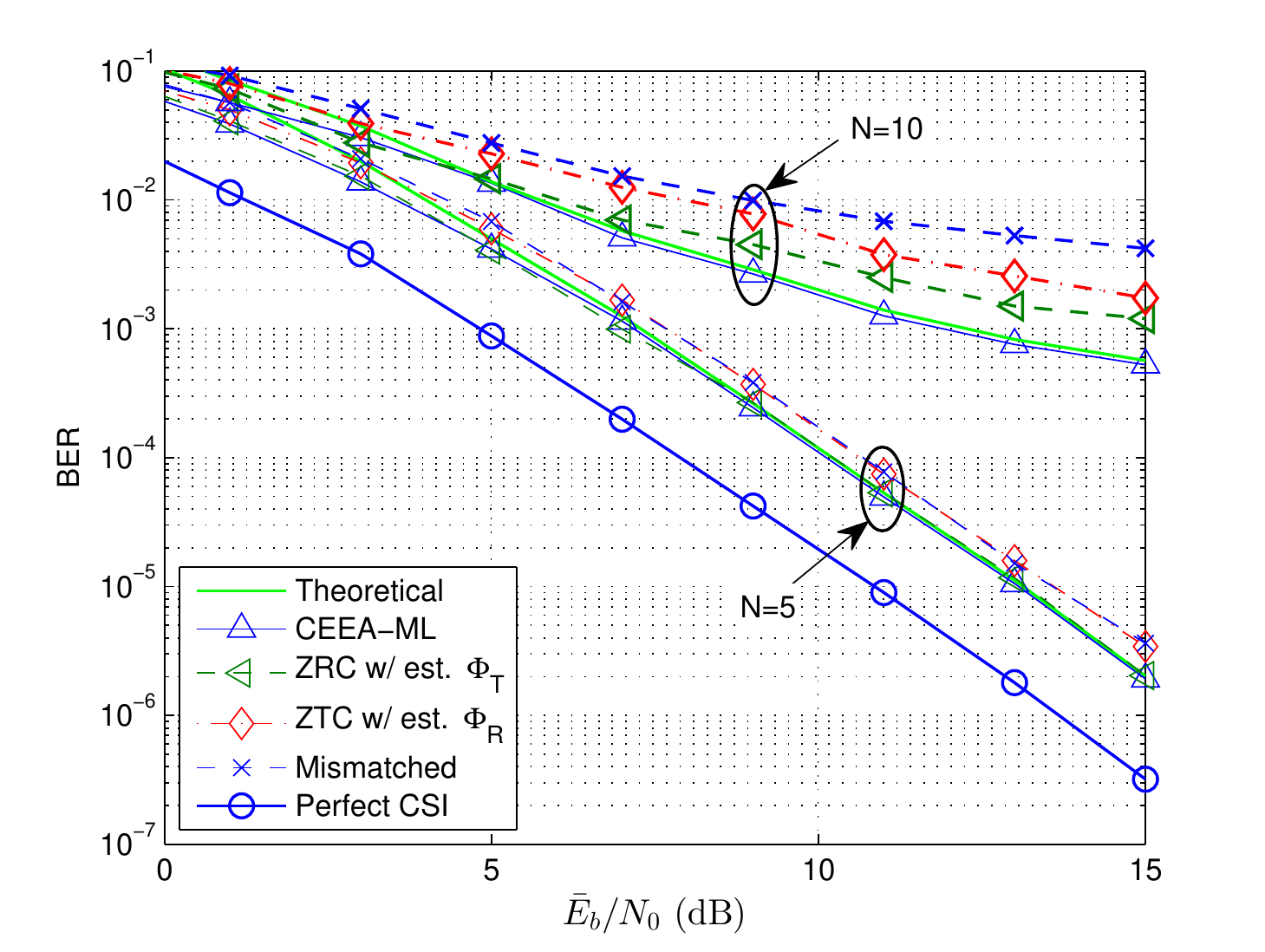}}}
        \caption{Effect of imperfect statistics on the MB-CE-aided ZRC and ZTC SM detectors in exponentially-correlated channels;
        $N=5$ or $10$, $M=4$, $N_T=N_R=B=4$, $f_DT_s=0.01$, and (a) $r=t=0.8$, (b) $r=t=0.5$.}
         \label{fig:MB_CorrEst_4x4_M4N5N10_Exp}
\end{figure}

%

\begin{figure}
  \centering
  \mbox{\subfloat[]{\label{subfig:MB_CorrEst_4x4_M4N5N10_lamb05} \includegraphics[width=3.6in]{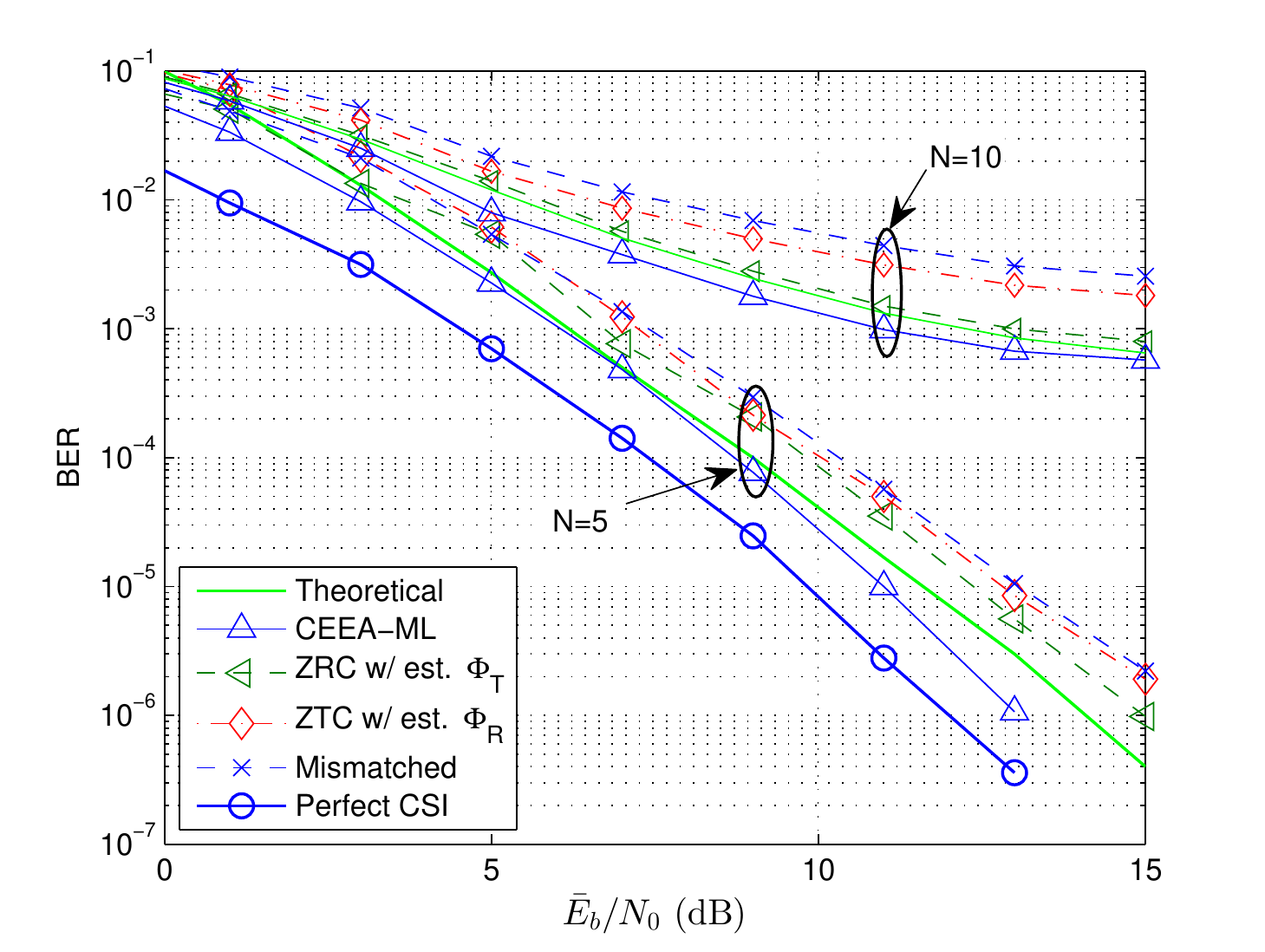}}}
  \mbox{\subfloat[]{\label{subfig:MB_CorrEst_4x4_M4N5N10_lamb1} \includegraphics[width=3.6in]{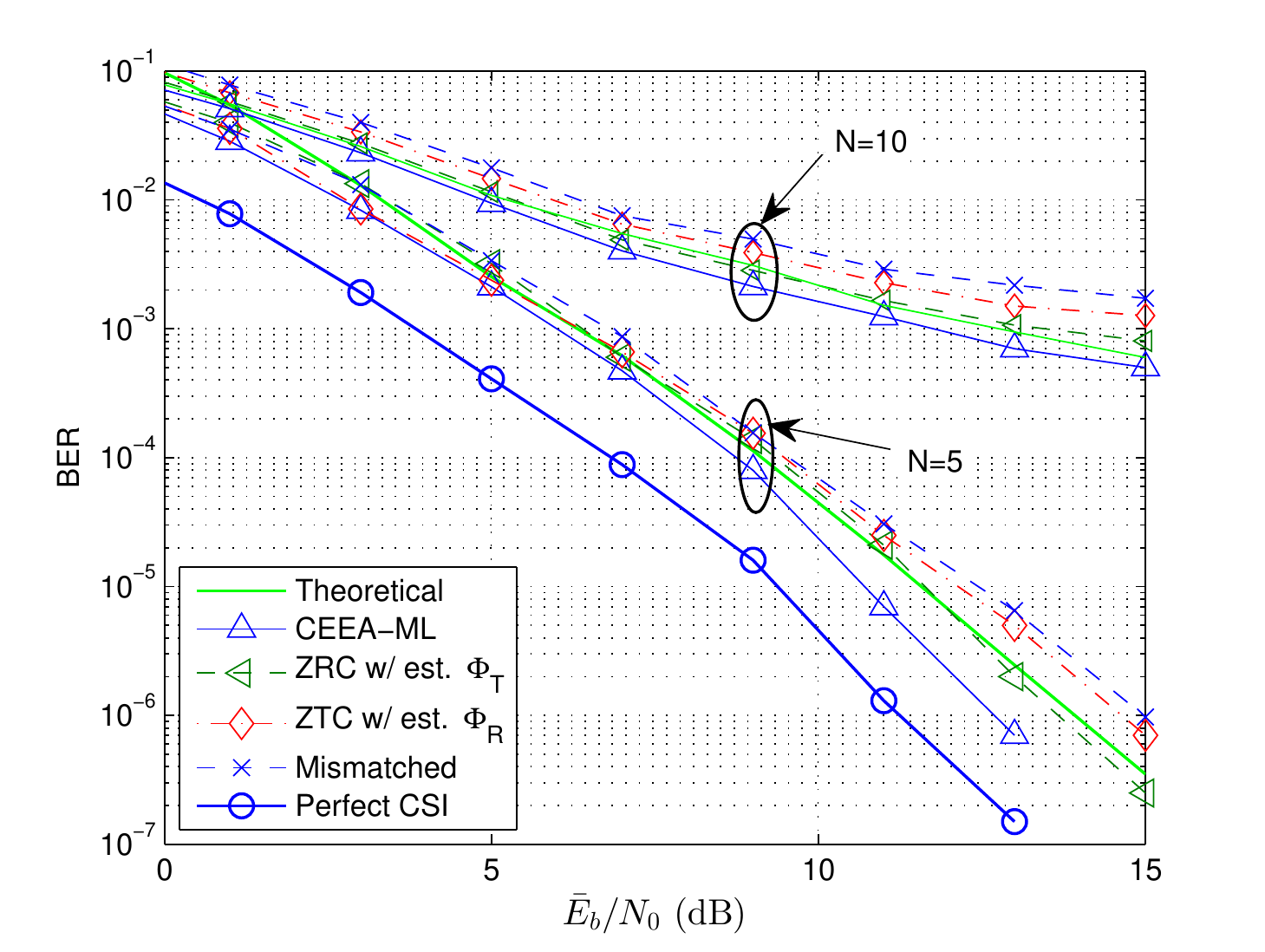}}}
        \caption{Effect of imperfect statistics on the MB-CE-aided ZRC and ZTC SM detectors in Bessel-correlated
        channels; $N=5$, $M=4$, $N_T=N_R=B=4$, $f_DT_s=0.01$, and (a) $\delta=0.5\lambda$; (b) $\delta=1\lambda$.}
         \label{fig:MB_CorrEst_4x4_M4N5N10_Bes}
\end{figure}
\begin{figure}
    \centering
        \includegraphics[width=3.6in]{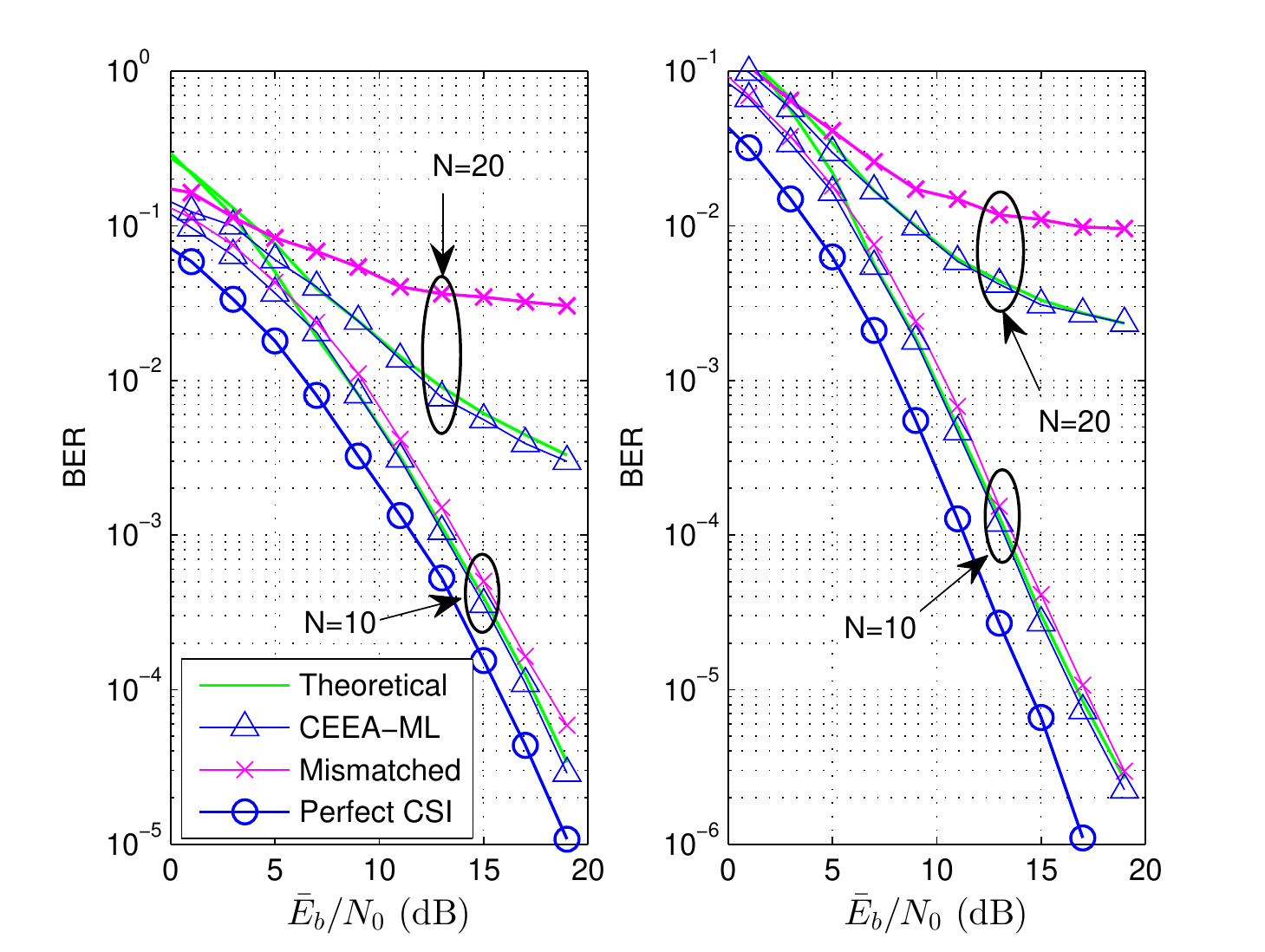}
        \caption{Performance of MB-CE-aided $16$-QAM SM systems ($5$-bit/transmission) with different frame sizes and
        spatial correlations; $N=10$ or $20$, $N_T=B=2$, $N_R=4$, $f_DT_s=0.01$, and (left) $r=t=0.8$; (right) $r=t=0.5$.}
\label{fig:MB_4x2_16QAM}
\end{figure}

The above results assume perfect statistical correlation information is available, we consider the impact of
imperfect spatial correlation information in Figs. \ref{fig:MB_CorrEst_4x4_M4N5N10_Exp} and \ref{fig:MB_CorrEst_4x4_M4N5N10_Bes}
where $\mb{\Phi}_R$ and $\mb{\Phi}_T$ used by {\it ZTC} and {\it ZRC} detectors are estimated by first taking the time averages over
three consecutive pilot blocks
\begin{subequations}
\begin{IEEEeqnarray}{rCl}
  \bar{\mb\Phi}_R&\defeq&\frac{1}{3N_T}\sum_{k=0,N,2N}
  \hat{\mb H}(k)\hat{\mb H}^H(k),\label{eq:SpCor_initialest1}\\
  \bar{\mb\Phi}_T&\defeq&\frac{1}{3N_R}\sum_{k=0,N,2N}
  \hat{\mb H}^T(k)\hat{\mb H}^*(k)
\label{eq:SpCor_initialest2}
\end{IEEEeqnarray}
\end{subequations}
and these initial estimates are then improved by (the temporal correlation is similarly estimated) $[\hat{\mb\Phi}_R]_{ij}=\hat{r}^{|i-j|}
\times\mathrm{sgn}(\Re\{[\bar{\mb\Phi}_R]_{ij}\})$ and $[\hat{\mb\Phi}_T]_{ij}=\hat{t}^{|i-j|}
\mathrm{sgn}(\Re\{[\bar{\mb\Phi}_T]_{ij}\})$, where
\begin{subequations}
\begin{IEEEeqnarray}{rCl}
  \hat{r}&=&\arg\min_{0\leq r<1}\sum_{i,j}\left|r^{|i-j|}-\left|[\bar{\mb\Phi}_R]_{ij}\right|\right|^2,\label{eq:r_hat}\\
  \hat{t}&=&\arg\min_{0\leq t<1}\sum_{i,j}\left|t^{|i-j|}-\left|[\bar{\mb\Phi}_T]_{ij}\right|\right|^2.\label{eq:t_hat}
\end{IEEEeqnarray}
\end{subequations}
The performance of ZTC and ZRC detectors shown in Figs. \ref{fig:MB_4x4_M4N5N10_Exp} and \ref{fig:MB_CorrEst_4x4_M4N5N10_Exp}
indicates that the refined spatial correlation estimator (\ref{eq:r_hat}) and (\ref{eq:t_hat}) gives fairly accurate estimations.
Both detectors keep their performance advantages over the {\it mismatched} counterparts when $N=10$. But with a denser pilot
$N=5$ (thus smaller mismatch error), the {\it ZTC detector} fails to offer noticeable gain due perhaps to additional approximation
(\ref{eq:ztcapp}) used. In Fig. \ref{fig:MB_CorrEst_4x4_M4N5N10_Bes}, the channel correlation follows (\ref{eq:SpCorr_Bessel})
but the receiver still assumes (\ref{eq:SpCorr_Exp}) and uses the estimator (\ref{eq:r_hat}) and (\ref{eq:t_hat}).
In spite of the correlation model discrepancy, the detectors still outperform the {\it mismatched} one. The theoretical performance
bound of the {\it CEEA-ML detector} analyzed in Section \ref{sec:theo} is also shown in Figs. \ref{fig:MB_CorrEst_4x4_M4N5N10_Exp}
and \ref{fig:MB_CorrEst_4x4_M4N5N10_Bes}. Except in lower $\bar{E}_b/N_0$ region, the theoretical bounds are tight and give reliable
numerical predictions. 
Similar accurate theoretical predictions and effect of $N$ are found in Fig. \ref{fig:MB_4x2_16QAM} for three $16$-QAM SM detectors.


\begin{figure}
    \centering
        \includegraphics[width=3.6in]{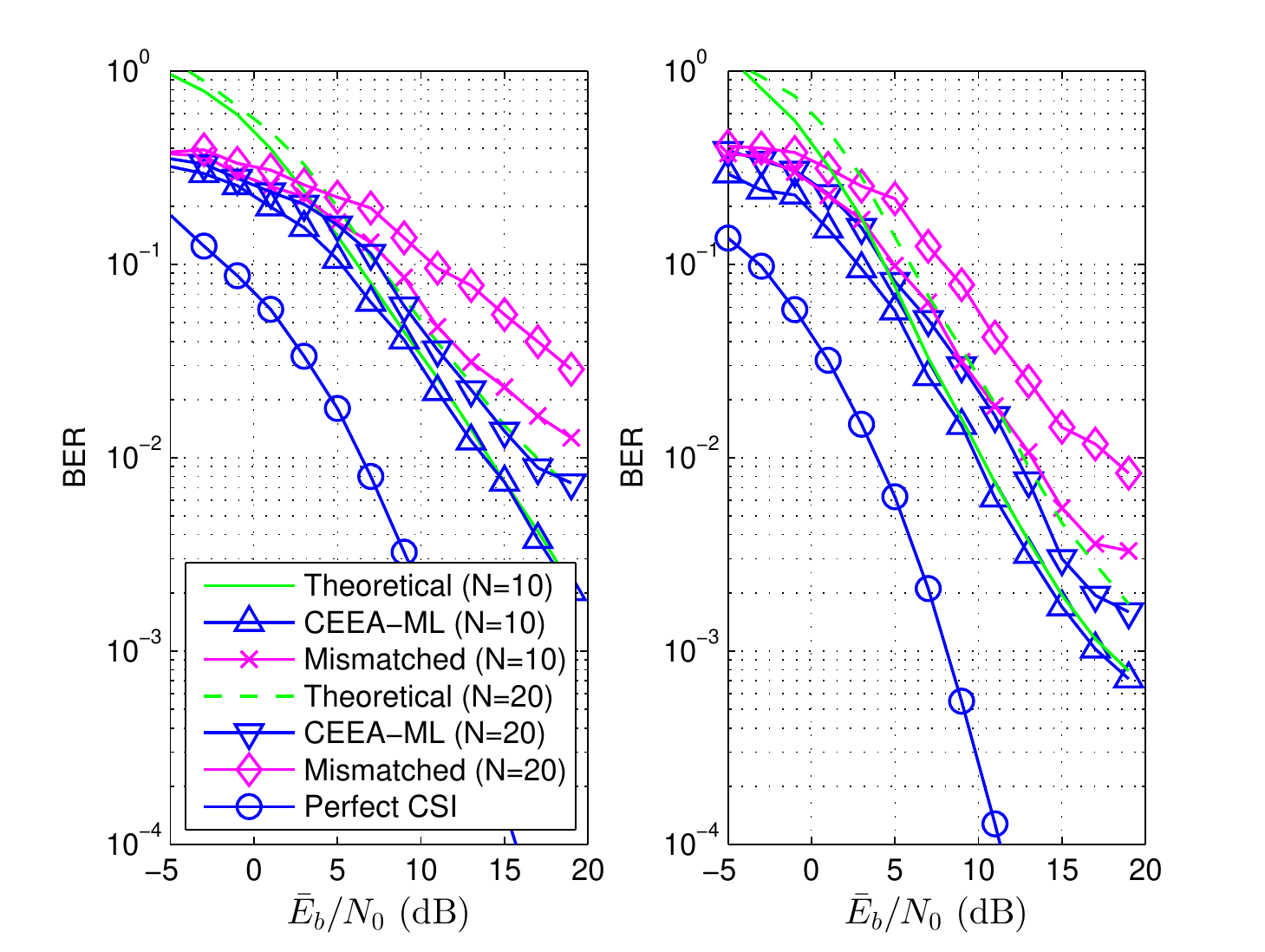}
        \caption{Performance of $5$-bit/transmission DD-CE-aided SM systems with
        different frame sizes and exponential spatial correlation; $16$-QAM $\mathcal{A}_M$,
        $N=10$ or $20$, $N_T=B=2$, $N_R=4$, $f_DT_s=0.01$, and (left) $r=t=0.8$; (right) $r=t=0.5$.}
        \label{fig:DD_4x4_M4N5}
\end{figure}

\begin{figure}
    \centering
        \includegraphics[width=3.6in]{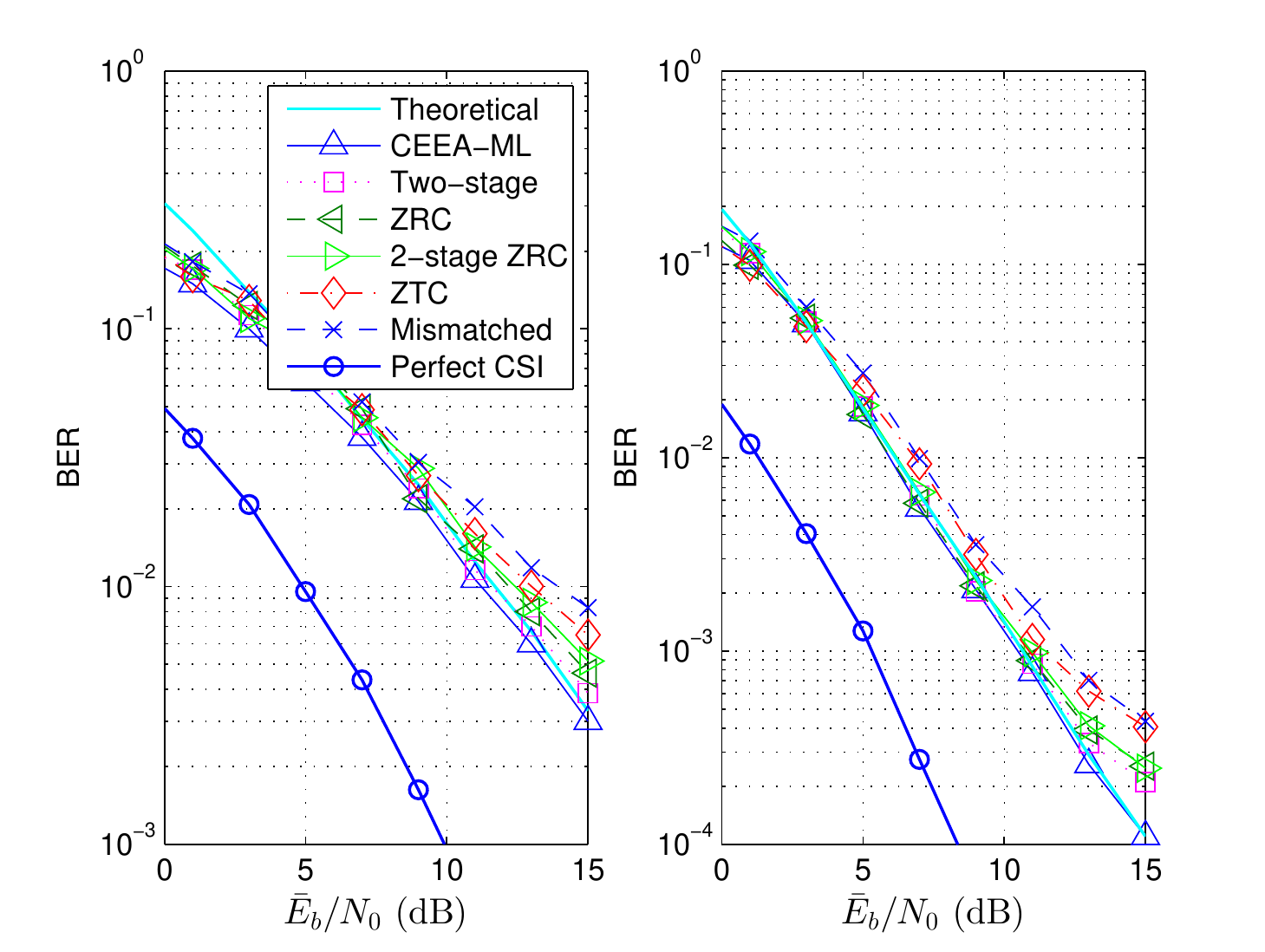}
        \caption{Performance of a $3$-bit/transmission DD-CE-aided SM system
        in exponentially-correlated channels; $N=10, M=4$, $N_T=B=2$,
        $N_R=4$, $f_DT_s=0.01$, and (left) $r=t=0.8$; (right) $r=t=0.5$.}
        \label{fig:DD_4x4_M4N10}
\end{figure}

\subsection{DD-CE-Aided SM Detectors}
Figs. \ref{fig:DD_4x4_M4N5} and \ref{fig:DD_4x4_M4N10} present the performance of the DD-CE-aided detectors.
As expected, the proposed detectors outperform the {\it mismatched} one. The effects of the pilot density,
correlation level and other behaviors of these detectors are similar to those observed in MB-CE-aided detectors.
But the DD-CE-aided detectors are more sensitive to the CSIR error. This is because the way the channel estimate
is updated, which is likely to be outdated in a fast fading environment and any detection error will propagate
until the next pilot block is received. The MB-CE-aided detectors which use three consecutive pilot blocks to
interpolate the time-varying channel response are less sensitive to time selectivity.

In Figs. \ref{fig:MB_4x2_16QAM} and \ref{fig:DD_4x4_M4N5}, we compare the effect of pilot density on both
classes of detectors. With long frame size ($N=20$), both MB- and DD-CE-aided detectors give unsatisfactory
performance although the former is slightly better. But if we increase the pilot density to $N=10$, the MB-CE-aided
detector offers more significant improvement: doubling the pilot density gives a $3.5$ dB gain at $\BER=1\times10^{-2}$
and $r=t=0.8$ (or $2.9$ dB at $r=t=0.5$) for the ML-MB detector, in contrast to the $2.6$ dB ($2.4$ dB) gain for
the ML-DD detector. Obviously, the CSI is of great importance and the {\it CEEA-ML detectors} significantly outperform
the {\it mismatched} ones when CSI is accurate. The DD-CE is improved with a smaller QAM constellation $\mathcal{A}_M$;
see Fig. \ref{fig:DD_4x4_M4N10} where $M=4$ is assumed. The {\it ZRC} and {\it ZTC detectors} ignore part of spatial
correlation, hence, it is only natural that their performance becomes closer to that of the {\it CEEA-ML detector}
as the spatial channel decorrelates, i.e., as $r$ and $t$ become smaller.

\begin{figure}
    \centering
        \includegraphics[width=3.6in]{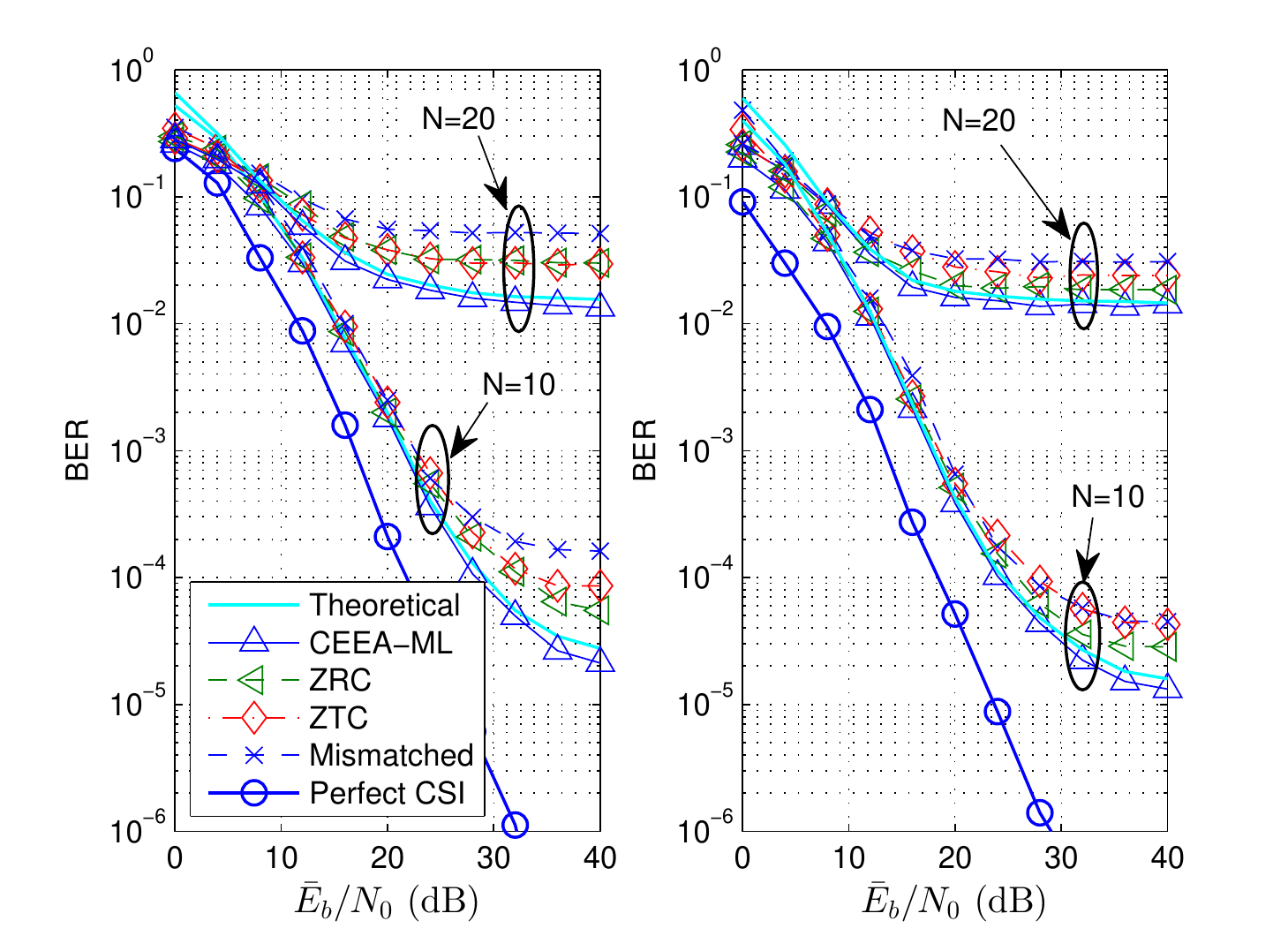}
        \caption{Performance of the MB-CE-aided SMX systems in exponentially-correlated
        channels; $N=10$ or $20$, $M=4$, $N_T=N_R=B=2$, $f_DT_s=0.01$, and (left) $r=t=0.8$; (right) $r=t=0.5$.}
        \label{fig:smx_MB_2x2_M4N10_r05r08}
\end{figure}

\subsection{CEEA-ML Detection of SMX Signals}
Finally in Fig. \ref{fig:smx_MB_2x2_M4N10_r05r08}, we show the performance of the SMX system in S-T
correlated channels using the {\it CEEA-ML detector} (\ref{eq:uni_ML_MB}). The system parameter values
are $B=N_{T}=N_{R}=2$ and $M=4$ so that it yields a rate of $4$ bits/transmission. The figure reveals
that the {\it CEEA-ML detector} and its suboptimal variations also bring about performance gain against
the {\it mismatched} one. The SM system with 4 bits/transmission and the same frame size, as shown in Fig.
\ref{fig:MB_CorrEst_4x4_M4N5N10_Exp}, achieves the same BER performance with significantly lower SNR.
The SMX systems result in higher BER error floors whereas the SM counterparts yield performance that is
much closer to that achieved with perfect CSI when frame size is $20T_s(N=10)$. This is because in the
high-SNR regime where the ICI is the dominant deteriorating factor for an SMX system, a high spatial
correlation may result in occasionally deep-fade across all spatial channels (all $|h_{ij}|$'s are small)
and a burst of erroneous symbols. Since the SM systems do not suffer from ICI, a rare single-channel fade
has less severe impact on its BER performance. Nevertheless, using imperfect CSI still helps to reduce
an SMX system's error floor. We present the MB-CE-aided SMX detectors' performance only as the DD-CE-aided
detectors give even worse performance.

\section{Conclusion}\label{sec:conclusion}
We have derived ML and various suboptimal detector structures for general MIMO (including SM and SMX)
systems that take into account practical design factors such as the channel's S-T correlations, the
channel estimator used and the corresponding estimation error. The pilot-assisted MB and DD channel
estimators we considered are simple yet efficient for estimating general S-T correlated MIMO channels.

The suboptimal detectors are obtained by simplifying the ML detector's exhaustive search, the spatial
correlation structure, the likelihood function, or a combination of these approximations. The complexities
of the ML, suboptimal and mismatched detectors are analyzed. The effects of space and/or time selectivity
and CSI error using MB or DD channel estimators on the system performance are studied via both analysis
and computer simulations. Their performance is compared with that of perfect CSIR detectors. We provide
numerical examples to verify the usefulness of our error rate analysis and to demonstrate how the CSI
uncertainty affects various detectors' BER performance and find when the fading channel's time or spatial
selectivity has to be taken into consideration. We also suggest a model-based spatial correlation estimator
that yields quite accurate estimates. The performance of spatial multiplexing system is studied as well.
The numerical results also enable us to find the channel conditions and performance requirements under which
the low-complexity suboptimal detectors incur only minor performance degradation and become viable
implementation choices.

\end{document}